\documentclass[pdflatex,sn-basic,iicol]{sn-jnl}

\usepackage{graphicx}%
\usepackage{adjustbox}%
\usepackage{multirow}%
\usepackage{amsmath,amssymb,amsfonts}%
\usepackage{amsthm}%
\usepackage{mathrsfs}%
\usepackage[title]{appendix}%
\usepackage{xcolor}%
\usepackage{textcomp}%
\usepackage{manyfoot}%
\usepackage{booktabs}%
\usepackage{algorithm}%
\usepackage{algorithmicx}%
\usepackage{algpseudocode}%
\usepackage{listings}%
\usepackage{subcaption}%

\theoremstyle{thmstyleone}%
\newtheorem{theorem}{Theorem}
\newtheorem{proposition}[theorem]{Proposition}%

\theoremstyle{thmstyletwo}%
\newtheorem{example}{Example}%
\newtheorem{remark}{Remark}%

\theoremstyle{thmstylethree}%
\newtheorem{definition}{Definition}%

\newcommand{\E}{\mathbb{E}}
\newcommand{\GP}{\text{GP}}
\newcommand{\mk}{\mathbf{k}} 
\newcommand{\mK}{\mathbf{K}}
\newcommand{\bfx}{\mathbf{x}}

\begin{document}

\title[Hierarchical additive interaction modelling with Gaussian process prior and its efficient implementation for  multidimensional grid data]{Hierarchical additive interaction modelling with Gaussian process prior and its efficient implementation for  multidimensional grid data}


\author*[1]{\fnm{Sahoko} \sur{Ishida}}\email{sahoko.ishida@ox.cs.ac.uk}

\author[2,3]{\fnm{Francesca} \sur{Panero}}\email{francesca.panero@uniroma1.it}

\author[3]{\fnm{Wicher} \sur{Bergsma}}\email{w.p.bergsma@lse.ac.uk}

\affil*[1]{\orgdiv{Department of Computer Science}, \orgname{University of Oxford}, \orgaddress{\street{Park Street}, \city{Oxford}, \postcode{OX1 3QG}, \country{United Kingdom}}}

\affil[2]{\orgdiv{Department of Methods and Models for Economics, Territory and Finance}, \orgname{Sapienza University of Rome}, \orgaddress{\street{Via del Castro Laurenziano 9}, \city{Rome}, \postcode{00161}, \country{Italy}}}

\affil[3]{\orgdiv{Department of Statistics}, \orgname{London School of Economics and Political Science}, \orgaddress{\street{Houghton Street}, \city{London}, \postcode{WC2A 2AE}, \country{United Kingdom}}}

\abstract{Additive Gaussian process (GP) models offer flexible tools for modelling complex non-linear relationships and interaction effects among covariates. While most studies have focused on predictive performance, relatively little attention has been given to identifying the underlying interaction structure, which may be of scientific interest in many applications. In practice, the use of additive GP models in this context has been limited by the cubic computational cost and quadratic storage requirements of GP inference.
This paper presents a fast hierarchical additive interaction GP model for multi-dimensional grid data. A hierarchical ANOVA decomposition kernel forms the foundation of our model, which incorporate main and interaction effects under the principle of marginality. Kernel centring ensures identifiability and provides a unique, interpretable decomposition of lower- and higher-order effects. For datasets forming a multi-dimensional grid, efficient implementation is achieved by exploiting the Kronecker product structure of the covariance matrix. Our contribution is to extend Kronecker-based computation to handle any interaction structure within the proposed class of hierarchical additive GP models, whereas previous methods were limited to separable or fully saturated cases. The benefits of the proposed approach are demonstrated through simulation studies and an application to high-frequency nitrogen dioxide concentration data in London.
}




\maketitle

\section{Introduction}\label{intro}
Gaussian process (GP) regression provides a flexible framework for modelling complex, nonlinear relationships in data. The flexibility of GPs arises primarily from the specification of the covariance function, or kernel, which encodes assumptions about smoothness and dependence. Different kernels can be combined through addition or multiplication, and both combinations preserve the validity of the covariance function. Additive kernels are often used to capture main effects, while multiplicative kernels can represent interactions among variables. The importance of modelling such interactions, and the trade-off between interpretability and predictive accuracy in flexible kernel design, was recognised early by \citet{plate1999accuracy}.

A substantial body of work has explored additive Gaussian process (GP) models as interpretable and flexible approaches for incorporating the additive effects of covariates and their interactions. Early contributions such as \citet{kaufman2010BayesianfANOVA} and \citet{Duvenaud_2011_NIPS_additiveGP} demonstrated that additive GPs can effectively decompose complex functions into sums of main and interaction components. \citet{duvenaud2013structure} presented automatic kernel structure discovery of relevant features and combinations. More recent developments, including those by \citet{cheng2019additive} and \citet{timonen2021lgpr}, have extended these ideas to longitudinal data and highlighted their value for explainable and interpretable modelling. \citet{lu2022additive} revisited the additive GP formulation of \citet{Duvenaud_2011_NIPS_additiveGP} and addressed identifiability issues by orthogonalising kernels. They further proposed the use of Sobolev indices to quantify the contribution of each component of the additive function. Their work is also closely connected to the functional ANOVA (fANOVA) \citep{wahba1990spline, huang1998projection, stone1994use} decomposition.

This paper follows the line of additive GP and functional ANOVA research but places particular emphasis on interaction modelling under the principle of marginality \citep{nelder1977reformulation}, which requires that higher-order interaction terms be accompanied by their relevant lower-order components. While many existing additive GP formulations adhere to this principle implicitly, they often fit models that include a larger set of interaction terms than are actually supported by the data. Such models may still achieve good predictive performance, but they can obscure interpretability and hinder the identification of which interaction effects are genuinely present.

To address this issue, we consider regression functions belonging to a hierarchical additive interaction class, where each component follows a GP prior defined through a hierarchical ANOVA decomposition kernel \citep{bergsma2023additive}. Although related ideas have appeared in the broader functional ANOVA and additive modelling literature, this formulation has not been clearly developed within the GP framework. Our proposed model formalises this structure and use centring of kernels when constructing the hierarchical ANOVA decomposition. Kernel centring plays an important role in ensuring identifiability and facilitating interpretation, as it allows the estimated main and interaction effects to represent deviations from lower-order terms in an intuitive way. This is similar to the orthogonalisation of kernel used in \citet{lu2022additive}, both aiming to achieve identifiability and enhance interpretability of additive components.

One of the central challenges in GP modelling lies in its computational cost, which scales cubically with the number of observations and quadratically in storage. A wide range of scalable GP approaches has been proposed (see \cite{liu2020gaussian} for a review), including sparse approximations, inducing point methods, and approaches that exploit special structure in the covariance matrix. Among these, Kronecker-based methods are particularly attractive for data observed on multi-dimensional grids. Such structures commonly occur when observations are collected across multiple input dimensions whose combinations form a Cartesian product, for example, measurements taken at multiple spatial locations over repeated time points. In these cases, the covariance matrix of the GP can often be expressed as a Kronecker product of as a Kronecker product of covariance matrices defined over each input dimension (e.g., spatial and temporal). This enables substantial reductions in computational and memory cost and allows for near-linear scaling in favourable cases , as shown in \citet{saatcci2012scalable, wilson2014fast, gilboa2013scaling, flaxman2015fast}.

However, existing Kronecker approaches are generally limited to a subset of GP models whose kernels are separable, meaning they can be expressed as a single tensor product. This class includes interaction-only or models that contain all possible interaction terms. While such models can be suitable for certain applications and can yield good predictive performance, they may not align with research questions in areas such as the social or medical sciences, where identifying which interaction effects are present is a key research question. For these settings, being able to efficiently estimate candidate models with different interaction structures is particularly valuable.

To address this gap, we extend the Kronecker framework to accommodate any hierarchical additive interaction model constructed using the proposed hierarchical ANOVA kernel, where the centring of kernels plays a crucial role. This extension allows for efficient computation without sacrificing the flexibility to include only the interaction structures supported by the data.

We assess the proposed method through two simulation studies: one examining its scalability to large datasets, and another evaluating its ability to identify the interaction effects underlying the data-generating process. Finally, we demonstrate the practical utility of the method through an application to hourly nitrogen dioxide ($\text{NO}_2$) concentration data in London, highlighting its interpretability and computational efficiency.

The remainder of the paper is structured as follows. Section~\ref{sec: additive GP} introduces the proposed additive GP model with a hierarchical ANOVA decomposition kernel for hierarchical additive interaction modelling. Section~\ref{sec: kronecker trick} presents our Kronecker-based approach for efficient implementation of the proposed model on multi-dimensional grid-structured data. Section~\ref{sec: simulation} reports results from two simulation studies, and Section~\ref{sec: application} illustrates the method through the real-world air quality application. Concluding remarks are provided in Section~\ref{sec: conclusion}.

\section{Hierarchical additive interaction Gaussian process model}\label{sec: additive GP}

Consider a regression model $y=f({x})+\epsilon$ for a real-valued response $y$ and  $P$ dimensional predictors $\bfx:=(x_1,x_2\ldots,x_P)$ where $x_j\in \mathcal{X}_j$ for $j\in \{1,\ldots,P\}$. 
Let $[P]:= \{1,\ldots,P\}$. An additive regression model takes the form
\begin{equation}
    f(\mathbf{x})=\sum_{t\in\mathcal{T}}f_t(\mathbf{x}_t)
\end{equation}
where $\mathcal{T}\subseteq 2^{[P]}$ is a subset of the power set of indices. 
Some common instances are the cases of $\mathcal{T}=\{\emptyset, \{1\},\dots,\{P\}\}$ and $\mathcal{T}=2^{[P]}$, corresponding respectively to an additive model with intercept and all the main effects:
\begin{equation}\label{eq: main additive function}
    f(\bfx) = f_\emptyset + f_1(x_1) + f_2(x_2) + \ldots + f_P(x_P)
\end{equation}
where $f_\emptyset$ is the constant term, and the saturated model:
\begin{align}\label{eq: saturated additive function}
    f(\bfx) =& f_\emptyset + \sum_{l=1}^P f_l(x_l) + \sum_{l_1=1}^P\sum_{l_2=l_1+1}^P f_{l_1 l_2}(x_{l_1},x_{l_2})\notag\\ 
    &+\dots+f_{12\ldots P}(x_1,x_2\ldots,x_P). 
\end{align}
which is the most complex and takes into account all the possible interaction terms.
In many applications, especially when $P$ is large, it is neither necessary nor desirable to include the full set of interaction terms in \ref{eq: saturated additive function}. Instead, we consider a structured subset of these terms that respects a hierarchical principle, or also known as principle of marginality \citep{nelder1977reformulation}, which ensurers that whenever a higher-order interaction is included, all of its lower-order component interactions, as well as all the main effects, are also present. The next section formalises this hierarchical additive–interaction class. Throughout the paper, we assume a mean zero prior on the regression function $f$ i.e., $f\sim \GP(0,k)$ where the kernel $k$ is defined on $\mathcal{X}=\mathcal{X}_1\times\mathcal{X}_2\times\ldots\times\mathcal{X}_p$. 

\subsection{Hierarchical ANOVA decomposition kernels}\label{sec: anova decomposition kernel}

In this paper, we will consider a specific class of additive regression functions that are defined over hierarchical families of indices $\mathcal{S}$. To proceed, we first give this definition.
 \begin{definition}
     Given a set of indices $[P]$, consider the tuple $(2^{[P]}, \subseteq)$ of the power set $2^{[P]}$ equipped with the inclusion operator $\subseteq$. We define a \textbf{hierarchical set} to be any $\mathcal{C}\subseteq 2^{[P]}$ such that for all $u \in \mathcal{C}$, $v\in\mathcal{C} \text{ for all } v\subseteq u$. We define a \textbf{hierarchical family of indices with singletons} $\mathcal{S}$ to be a set given by the union of the singleton elements \{1\},\dots, \{P\}, the empty set $\emptyset$, and any user-selected hierarchical set $\mathcal{C}\subseteq 2^{[P]}$. 
 \end{definition}

For example, with $P=4$ the set $\mathcal{S}=\{\emptyset, \{1\},\dots,\{4\}, \{1, 2\}, \{1, 3\}\}$ is hierarchical with singletons, because it is the union of the empty set, the singletons and the hierarchical sets $\{\{1\}, \{2\}, \{1, 2\}\}$ and $\{\{1\}, \{3\}, \{1, 3\}\}$. Instead, the set $\mathcal{S}=\{\emptyset, \{1\},\dots,\{4\}, \{1, 2, 3\}\}$ does not satisfy the definition because $\{1,2\}, \{1,3\}, \{2,3\}$ are included into $\{1, 2, 3\}$ but do not belong to $\mathcal{S}$.
Note that the inclusion of the empty set and the singletons implies that in our models we will always have an intercept and main effects terms. For simplicity, since we will always focus on hierarchical families with singletons, we will equivalently call them hierarchical families or sets.\\



We are now ready to introduce a hierarchical ANOVA decomposition kernel, which will be used as the kernel of the Gaussian process prior in our model.
\begin{definition}\label{hierarchical_additive_kernel}
    Given a hierarchical family of indices $\mathcal{S}\subseteq 2^{[P]}$, a \textbf{hierarchical additive interaction kernel} is a kernel that satisfies
     \begin{equation}\label{eq: hierarchical additive kernel}
         k(\bfx,\bfx') = \sum_{u\in\mathcal{S}} k_u(\bfx_u,\bfx'_u),
    \end{equation}
 where $\bfx_u := \left(x_j\right)_{j\in u}$ and $k_u$ is a kernel over the general space $\mathcal{X}_u$ where $\bfx_u$ live.
\end{definition}

We can model a regression function of the form
 \begin{equation}\label{eq: decomposed function}
    f(\bfx) = \sum_{u\in\mathcal{S}}f_u(\bfx_u) 
 \end{equation}
by using the kernel in \eqref{eq: hierarchical additive kernel} as prior specification for the covariance function of a Gaussian process regression.\\
 
The simplest instance of hierarchical set is $\mathcal{S}=\emptyset \cup \{\{l\}:l\in[P]\}$, whose corresponding hierarchical additive kernel is 
\begin{equation}\label{eq: main additive kernel}
    k(\bfx,\bfx') = k_\emptyset(\bfx,\bfx') + \sum_{l=1}^P k_l(x_l, x_l')
\end{equation}
where $k_\emptyset(\bfx,\bfx')=c\in\mathbb{R}$ for any $\bfx, \bfx'$ is the constant kernel. Without loss of generality, from now on we will take $c=1$. The function $k_l:\mathcal{X}_l\times\mathcal{X}_l\rightarrow\mathbb{R}$ are the base kernel for variable $x_l$. We identify this kernel the \textit{main effect kernel}, since it can be used as a convenient prior for the additive main effects model of \eqref{eq: main additive function}.
 
At the other end of the spectrum, corresponding to $\mathcal{S}=2^{[P]}$, the most complete kernel will generate the saturated interaction model, where all possible interactions are included:
\begin{equation}\label{eq: saturated interaction kernel}
    k(\bfx,\bfx') = 1 + \sum_{u\in 2^{[P]}} k_u(\bfx_u, \bfx_u').
\end{equation}
Although the interactions do not need any specific structure in their most general form, in the rest of the work we will focus on interaction kernels given by the tensor product of the main effect kernels, for which we state a separate definition. 

\begin{definition}\label{hierarchical_ANOVA_kernel}
    Given a hierarchical family of indices $\mathcal{S}\subseteq 2^{[P]}$ and a set of main-effects kernels $k_l$ over $\mathcal{X}_l$ for each $l=1,\dots,P$, a \textbf{hierarchical ANOVA interaction kernel} is a hierarchical additive interaction kernel as defined in \eqref{eq: hierarchical additive kernel} such that, for all $u\in\mathcal{S}$ with $|u|\ge 2$, the kernels associated with $u$ satisfy the following factorisation:
    \begin{equation}\label{eq: tensor product kernel}
        k_u(\bfx_u,\bfx'_u) = \otimes_{l\in u}k_l(x_l,x'_l)
    \end{equation}
    where $\otimes$ is the tensor product over the space $\prod_{l\in u} \mathcal{X}_l$.
\end{definition}

This tensor-product specification offers the advantage that, with appropriate centring of the base kernels, the resulting components automatically satisfy the functional ANOVA constraints, which will be detailed in Section \ref{sec: centring}. With this construction, the prior for the overall function $f$ in the saturated model of \eqref{eq: saturated additive function} is a zero-mean GP with kernel
\begin{equation}\label{eq: saturated ANOVA kernel}
    k(\bfx,\bfx') = \otimes_{l=1}^P \left(1+k_l(x_l,x'_l)\right).
\end{equation}
This is known in the literature as the ANOVA decomposition kernel, or more simply ANOVA kernel \citep{stitson1999support, durrande2013anova}. In this paper, we refer to this as \textit{saturated ANOVA decomposition kernel}. We will keep the tensor product structure of the kernel for all hierarchical models that contain interaction effects, not necessarily the saturated model.\\

\begin{figure}[]
    \centering
        \label{fig: main term anova kernel}
        \includegraphics[width=0.49\linewidth]{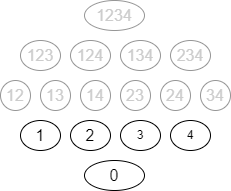}
        \includegraphics[width=0.49\linewidth]{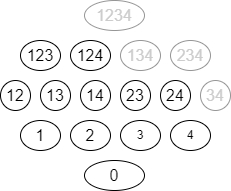}
        \vskip 3mm
        \includegraphics[width=0.49\linewidth]
        {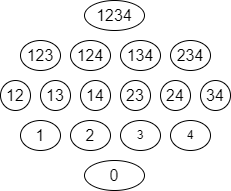}
        \includegraphics[width=0.49\linewidth]{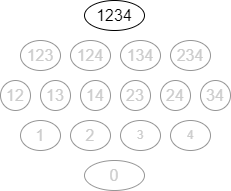}
        \caption{Visualisation of ANOVA decomposition kernels and a tensor product kernel with $P=4$ dimensional covariates. From top left to bottom right, we have main effect, hierarchical, saturated and tensor product kernels. The term $0$ in the panels refers to the constant term, and the terms $1$ to $4$ refers to main effect term that corresponds to $k_l$ for $l=1,\ldots,4$. The remaining terms are interaction effect terms, e.g., the term $123$ models the three-way interaction effect involving the covariates $\mathbf{x}_1$,$\mathbf{x}_2$ and $\mathbf{x}_3$. Panel (b) is an example of a hierarchical ANOVA decomposition kernel. Adding the term $34$ to this example gives another example of such kernel. If we are to include the term $134$ and/or $234$, the term $34$ should also be added in order to ensure a hierarchical structure.} 
        \label{fig: kernel structure example}
\end{figure}

Figure \ref{fig: kernel structure example} illustrates the differences between the interaction structures discussed in this section with $P=4$. It is worth noting that there are many hierarchical ANOVA kernels, each corresponding to a different hierarchical family of indices $\mathcal{S}$, and figure \ref{fig: kernel structure example} shows one such example.

\subsection{Posterior}
Let the sample of observations be $\mathcal{D}=\{(\bfx_i,y_i)\}_{i=1}^n$. Given a regression function $f$ as in \eqref{eq: decomposed function} with an additional additive noise $\epsilon\sim N(0,\sigma^2)$, we specify a prior over it using a Gaussian process with zero mean function and a covariance kernel $k$ given by a hierarchical ANOVA decomposition kernel of def. \ref{hierarchical_ANOVA_kernel}, $\text{GP}(0,k)$. 

Under this specification, the posterior of $f$ given the data $\mathcal{D}$ is also a Gaussian process with mean function $\bar{m}(\bfx)$ and covariance function $\bar{k}(\bfx, \bfx')$, which we can write in closed form.
Call $\mathbf{K}_l \in \mathbb{R}^{n\times n}$ the Gram matrix of component $l\in[P]$ whose elements are $(\mathbf{K}_l)_{ij}=k_l(x_{il},x_{jl}')$. Due to the tensor product structure, we have $\mathbf{K}_u=\odot_{l\in u} \mathbf{K}_l$ with $\odot$ element-wise product, and $\mathbf{K} = \sum_{u\in\mathcal{S}}\mathbf{K}_u$.
The posterior mean and covariance for the term $u\in\mathcal{S}$ are
\begin{align}
    \bar{m}_u(\bfx_u) =& \mathbf{k}_u(\bfx_u)^\top(\mathbf{K}+\sigma^2\mathbf{I})^{-1}\mathbf{y}\label{eq: post mean decomposition}\\
    \bar{k}_u(\bfx_u,\bfx'_u) =& k_u(\bfx_u,\bfx'_u) \notag\\
    &-  \mathbf{k}_u(\bfx_u)^\top(\mathbf{K}+\sigma^2\mathbf{I})^{-1}\mathbf{k}_u(\bfx'_u)  \label{eq: post kernel decomposition}
\end{align}
where $\mk_u(\bfx_u) =(k_u(\bfx_u,\bfx_{1u}),\ldots, k_u(\bfx_{u},\bfx_{nu}))^\top$ and $\mathbf{y}= (y_1,\dots,y_n)^\top$.  Finally, the posterior mean of the function $f$ given $\mathcal{D}$ has the same additive structure
\begin{equation}\label{eq: overall posterior mean}
    \bar{m}(\bfx) = \sum_{u\in \mathcal{S}}\bar{m}_u(\bfx_u),
\end{equation}
while the kernel decomposes additively as
\begin{equation}\label{eq: overall posterior kernel}
    \bar{k}(\bfx,\bfx') = \sum_{u\in\mathcal{S}}\bar{k}_u(\bfx_u,\bfx_u') + \sum_{u\in\mathcal{S}}\sum_{v\in\mathcal{S}\backslash \{u\}}\bar{k}_{uv}(\bfx_u,\bfx'_v)\notag
\end{equation}
with
\begin{equation}\label{eq: post cross kernel}
    \bar{k}_{uv}(\bfx_u,\bfx'_v) = -\mk_u(\bfx_u)^\top(\mathbf{K}+\sigma^2\mathbf{I})^{-1}\mk_v(\bfx'_v).
\end{equation}
for $u\neq v$.
This cross covariance generally does not vanish unless some special independence structures hold. 

\subsection{Centring, identifiability and interpretation}\label{sec: centring}
The decomposition of \eqref{eq: decomposed function} is generally not unique, as discussed in \cite{ginsbourger2016anova, martens2019decomposing, lu2022additive}. For instance, we can subtract any constant from a component $f_u(\bfx_u)$ and add it to another term $f_{v}(\bfx_{v})$, $v\neq u$, and we would obtain the same function $f$. To make the decomposition unique and thus identifiable, it is standard to impose orthogonality constraints. For the family of probability measures $\{\nu_l\}_{l=1}^P$ on $\mathcal{X}_1,\dots,\mathcal{X}_P$ that originate the data, we require
\begin{equation}\label{eq: ANOVA constraint for identifiability}
    \int_{\mathcal{X}_l} f_u(\bfx_u) d\nu_l (x_l) = 0, \quad \forall l \in u, \text{ } u\in\mathcal{S}. 
\end{equation}
If each $f_u$ in the prior satisfies this constraint, the posterior decomposition is unique. 

In the GP setting with the tensor-product construction for $k_u = \otimes_{l\in u}k_l$, this can be achieved by replacing each base kernel $k_j$ with its centred version
\begin{align}\label{eq: centred kernel}
    k^{c}_l(x_l,x'_l) =& k_l(x_l,x'_l) - \mathbb{E}_{X_l'\sim \nu_l}[k_l(x_l,X_l')] \notag\\
    &- \mathbb{E}_{X_l\sim \nu_l}[k_l(X_l,x'_l)]\notag\\ 
    &+ \mathbb{E}_{(X_l,X'_l)\sim\nu_l}[k_l(X_l,X'_l)].
\end{align}
We refer to \ref{apx: kernel constraints} to show why this operation implies \eqref{eq: ANOVA constraint for identifiability}.

In practice, these expectations are often replaced by empirical averages over the observed inputs$\{x_{il}\}_{i=1}^n$, yielding the empirically centred kernel
\begin{align}\label{eq: empirically centred kernel}
    \tilde{k}^{c}_l(x_l,x'_l) =& k_l(x_l,x'_l) - \frac{1}{n}\sum_{i=1}^n k_l(x_l,x_{il}) \notag\\
    &- \frac{1}{n}\sum_{i=1}^n k_l(x'_l,x_{il}) \notag\\
    &+ \frac{1}{n^2}\sum_{i=1}^n\sum_{j=1}^n k_l(x_{il},x_{jl}).
\end{align}
This empirical construction ensures 
that the corresponding GP components have zero empirical mean:
$\sum_{i=1}^n f_l(x_{il})=0$. Applying \eqref{eq: empirically centred kernel} to Gram matrix $\mathbf{K}_l$ where $\{\mathbf{K}_l\}_{ij} = k(\bfx_i,\bfx_j)$ is equivalent to 
\begin{equation}\label{eq: empirically centred Gram matrix}
    \tilde\mK^{(c)}_l = \mathbf{C}\mK_l\mathbf{C}, \quad 
    \mathbf{C} = \mathbf{I}_n -\frac{1}{n}\mathbf{1}_n\mathbf{1}_n^{\top}
\end{equation}
where $\mathbf{1}_n$ is the $n$-dimensional vector of ones.

Imposing the centring constraints \eqref{eq: ANOVA constraint for identifiability} ensures that each $f_u$ captures only the $|u|$-way interaction effect among the variables indexed by $u$ with all lower-order effects separated into their corresponding terms. This yields a unique, orthogonal decomposition in which main effects and interactions are clearly separated. For example, a main-effect term $f_j(x_j)$ can be interpreted as as the isolated effect of variable $x_j$ averaged over all other variables, while the constant term $f_{\emptyset}$ corresponds to the overall mean of the response. In the GP setting, if the prior terms satisfy the constraints, the posterior mean functions $\bar{m}_u$ inherit the same property. With empirical centring, $\sum_{i=1}^n \bar{m}(x_{il},x_{i,-l})=0$ for each $l \in u$, which indicates that all terms in additive regression function, including main effects and lower-order interaction effects have intuitive interpretation and can be understood as the effect averaged over the input. See Section \ref{sec: application} for
illustration with real-world data.


\subsection{Hyper-parameter estimation and model comparison}
Hyper-parameters in the GP model, including kernel parameters (denoted by $\theta$) and the noise variance $\sigma^2$, can be estimated either via empirical Bayes, by maximising the marginal log-likelihood, or via Markov chain Monte Carlo (MCMC) methods, such as Hamiltonian Monte Carlo (HMC). In both cases, evaluation of the marginal log-likelihood 
\begin{align}\label{eq: log  marginal likelihood}
\log{p(\mathbf{y}|\mathbf{x},\mathbf{\theta},\sigma^2)}  =&  -\frac{1}{2}\mathbf{y}^\top(\mK_\mathbf{\theta}+\sigma^2\mathbf{I}_n)\mathbf{y} \notag\\
&-\frac{1}{2}\log|\mK_\mathbf{\theta}+\sigma^2\mathbf{I}_n| - \frac{n}{2}\log 2\pi
\end{align}
is required, either for direct optimisation in empirical Bayes or at each iteration in MCMC sampling. 

For model comparison, a range of criteria can be applied. In our MCMC-based implementation with \texttt{Stan} \citep{stan},  we use Bayesian leave-one-out (LOO) estimate of out-of-sample predictive density \citep{vehtari2016bayesianloo, vehtari2016practical, yao2018stacking}, computed using the R package \texttt{loo} \citep{vehtari2024loopackage}, the marginal likelihood estimated via \texttt{bridgesampling}\citep{gronau2020bridge}, the deviance information criterion (DIC), and best-fit predictive density $\log p(\mathbf{y}|\bfx, \hat{\theta},\hat{\sigma}^2)$ where $(\hat{\theta},\hat{\sigma}^2)$ are the posterior mean. In our simulation study, we  computed DIC and best-fit predictive density.
\subsection{Related work}
The literature on ANOVA decompositions and additive Gaussian processes is extensive; here we briefly review strands most relevant to our work.
Our work is rooted in the long-standing literature on functional ANOVA (fANOVA) decompositions, which provide a systematic way to express multivariate functions as sums of main effects and interaction terms. This framework has been applied across a wide range of statistical and machine learning models: for instance, in splines (see \cite{gu2002smoothing}, for an overview), in tree-based methods for interaction detection \citep{lengerich2020purifying}, and in the development of Gaussian process ANOVA models (e.g. \cite{kaufman2010BayesianfANOVA}). More recently, ANOVA-inspired ideas have been adapted to additive Gaussian processes \citep{lu2022additive}, neural decomposition models that extend fANOVA to deep learning \citep{pmlr-v108-martens20a}, and to the I-prior framework \citep{bergsma2020regression} for functional priors in reproducing kernel Hilbert space \citep{bergsma2023additive}.

Also close to our setting are works on additive GP models, most notably \cite{Duvenaud_2011_NIPS_additiveGP} and \cite{lu2022additive}. The former propose an additive GP without ANOVA constraints, while the latter ensure identifiability of each component. Our approach differs in several ways. First, we allow for broader interaction structures: their construction specifies an interaction order $d\in[P]$ including all terms up to order 
$d$, which is a special case of the hierarchical additive–interaction class we introduce. Second, in terms of kernel constraints, we employ double centring rather than orthogonal projection (see Appendix \ref{apx: kernel constraints}). Third, regarding parametrisation, while all methods construct priors for interaction terms via tensor products of base kernels, in \cite{lu2022additive} all $d-$th order terms share a common scale parameter, whereas our formulation treats scaling more flexibly and does not increase the number of parameters to estimate, even when we add higher order interactions.
Despite these differences, the computational strategy we develop in the next section, which extends the Kronecker trick to GP regression with hierarchical ANOVA kernels, can be applied to the models presented in \cite{lu2022additive}.

\section{Efficient implementation using Kronecker product structure}\label{sec: kronecker trick}
A major computational bottleneck in GP regression arises from the need to evaluate the marginal likelihood in \eqref{eq: log  marginal likelihood} and posterior quantities in \eqref{eq: post mean decomposition}, \eqref{eq: post kernel decomposition} and \eqref{eq: post cross kernel}. 
Suppose we have observations $\{(\bfx_i, y_i)\}_{i=1}^n$, where each input takes the form $\bfx_i = (x_{i1}, \ldots, x_{iP})$ with $x_{il} \in \mathcal{X}_l$ for $l \in [P]$. 
Inference requires repeated operations involving the $n \times n$ covariance matrix $\mathbf{K} + \sigma^2 \mathbf{I}$. The main computational costs are (i) solving linear systems $(\mathbf{K} + \sigma^2 \mathbf{I})^{-1}\mathbf{v}$, which are needed for evaluating the log marginal likelihood as well as the posterior mean and covariance, and (ii) computing the log-determinant $\log |\mathbf{K} + \sigma^2 \mathbf{I}|$ which appears in the log likelihood. Both scale cubically in $n$, with $O(n^3)$ time and $O(n^2)$ storage complexity when performed naively.

These costs can be greatly reduced when the input points are observed on a multi-dimensional grid. Formally, we say that the design has a grid structure if the input space can be written as a Cartesian product $\tilde{\mathcal{X}} = \tilde{\mathcal{X}}_1 \times \tilde{\mathcal{X}}_2\times\ldots\times\tilde{\mathcal{X}}_P$ where each $\tilde{\mathcal{X}}_l$ is a finite set of observed values along the $l$-th coordinate. Let $n_l=|\tilde{\mathcal{X}}_l|$ denote its cardinality. Then the full design consists of all possible tuples $\bfx = (x_1,\ldots,x_P)\in\tilde{\mathcal{X}}$ and the total number of observations is $n = \prod_{l=1}^P n_l$. This Cartesian product structure induces a Kronecker product form in the covariance matrix, which allows efficient computation of both matrix–vector products and log-determinants via eigendecomposition of lower-dimensional factors.\\

Such Kronecker methods are most commonly exploited for GP models whose kernels are tensor products across input dimensions, corresponding to models that include only the highest-order interaction term. Our contribution is to show that this idea can be generalised to a broader class: additive-interaction GP models with hierarchical ANOVA kernels under centring constraints. In what follows, we first outline the general computational procedures underlying the Kronecker approach, before demonstrating how they can be applied to saturated and non-saturated hierarchical ANOVA kernels.

\subsection{General procedure of Kronecker approach}\label{sec: Kronecker general procedure}
To identify a Kronecker product involving $P$ matrices, we write $\bigotimes_{l=1}^P \mathbf{M}_l = \mathbf{M}_1 \otimes \mathbf{M}_2\otimes\ldots\otimes\mathbf{M}_P$. For $P-$dimensional grid data, the main idea of Kronecker methods is to decompose the Gram matrix in the form: 
\begin{equation}\label{eq: eigendecomposition of gram matrix example}
    \mathbf{K} = \left(\bigotimes_{l=1}^P {\mathbf{Q}}_l \right)\mathbf{D} \left(\bigotimes_{l=1}^P {\mathbf{Q}}_l \right)^\top 
\end{equation}
where each $\mathbf{Q}_l$, and hence also $\mathbf{Q} := \bigotimes_{l=1}^P {\mathbf{Q}}_l$, is orthonormal and $\mathbf{D}$ is diagonal with non-negative entries. 

Once we obtain this decomposition, the log determinant of the marginal covariance matrix can be computed by
\begin{equation}\label{eq: kronecker log determinant}
    \log|\mathbf{K}+\sigma^2\mathbf{I}| = \sum_{i =1}^n \log(\mathbf{D}_{i,i} +\sigma^2) 
\end{equation}
where $\mathbf{D}_{i,i}$ is the $i$-th diagonal element of $\mathbf{D}$. This costs $O(n)$ operations. 

The multiplication of the inverted matrix and a vector $\mathbf{v}$ can be expressed as 
  \begin{align}\label{eq: kronecker linear system}
      &\left(\mathbf{K}  + \sigma^2 \mathbf{I}_n\right)^{-1}\mathbf{v} =\notag\\ &\left(\bigotimes_{l=1}^P {\mathbf{Q}}_l \right)\left(\mathbf{D} + \sigma^2 \mathbf{I}_n \right)^{-1} \left(\bigotimes_{l=1}^P {\mathbf{Q}}_l \right)^\top\mathbf{v}.
  \end{align}
The inversion of the middle diagonal matrix can be done by simply inverting its diagonal elements. Evaluating the above also requires matrix-vector multiplication $(\bigotimes_{l=1}^P {\mathbf{Q}}_l)^\top\mathbf{v} = \left(\bigotimes_{l=1}^{P} {\mathbf{Q}}_l^\top\right) \mathbf{v}$. 
 We write $\mathbf{v}_P = \text{vec}(\mathbf{V}^\top\mathbf{Q}_P)$ with $\text{vec}(\mathbf{A})$ being a vectorisation operator transforming a $p\times q$ matrix $\mathbf{A}$ to a vector of length $pq$ by stacking the columns of the matrix, and $\mathbf{V}$ is a $n_P\times \frac{n}{n_P}$ matrix whose elements are filled with elements of vector $\mathbf{v}$ in column-major order. Computing $\mathbf{v}_P$ takes $O(n_l^2 \frac{n}{n_l}) = O(n n_l)$. Iteratively applying this to get $\mathbf{v}_{P-1} = \text{vec}(\mathbf{V}_P^\top\mathbf{Q}_{P-1}),\mathbf{v}_{P-2}=\text{vec}(\mathbf{V}_{P-1}^\top\mathbf{Q}_{P-2}),\ldots \mathbf{v}_1 =\text{vec}(\mathbf{V}_2^\top\mathbf{Q}_1)$ thus requires $O(n\sum_{l=1}^P n_l)$ operations, and the final vector $\mathbf{v}_1$ equals $(\bigotimes_{l=1}^P {\mathbf{Q}}_l)^\top\mathbf{v}$. The complete algorithm is described in \citet[Chapter 5]{saatcci2012scalable} and \citet{wilson2014fast}.\\

The Kronecker method has been used and proven useful for efficient implementation of GP models (e.g. \citet[Chapter 5]{saatcci2012scalable}, \citet{groot2014fast} \citet{wilson2014fast} \citet{flaxman2015fast}). However, the existing method is applicable to limited sub-models with so-called separable kernel structures. 
The literature focused on applying it the tensor-product kernel, but, as mentioned previously, using this kernel implies including only the interaction term of the highest order. This may be problematic in many applications where assessing the effect of each predictor is needed. 
We will start \ref{sec: kronecker hierarchical} by first showing how to  derive the decomposition for the saturated ANOVA decomposition kernel, which is an easy although necessary extension of the tensor-product case.
Using the saturated ANOVA kernel, though, means that we assume a saturated model, which could often overfit the data. In light of this, the main methodological contribution of the paper will be to prove that a Kronecker decomposition can also be found for the non trivial case of the more general, hierarchical ANOVA kernel.




\subsection{Gram matrix decomposition for hierarchical ANOVA kernels}\label{sec: kronecker hierarchical}

\subsubsection{The saturated case}

Let us assume that we have $P$-dimensional grid structure in the predictors and a response vector $\mathbf{y}$ of length $n$. If we use the saturated ANOVA decomposition kernel \eqref{eq: saturated ANOVA kernel}, the Gram matrix can be written as 
  \begin{equation*}
      \mathbf{K} = \bigotimes_{l=1}^P \Tilde{\mathbf{K}}_l 
  \end{equation*}
where $\Tilde{\mathbf{K}}_l = (\mathbf{1}_{n}\mathbf{1}_{n_l}^\top + \mathbf{K}_l)$ the Gram matrix of the $l-$th main effect term, and $\mathbf{K}_l= \{k_{l (ij)}\}_{n_l\times n_l}$ with $k_{l(ij)} = k_l(x_{il},x_{jl})$. Using the eigendecomposition, $\Tilde{\mathbf{K}}_l = \Tilde{\mathbf{Q}}_l \Tilde{\boldsymbol{\Lambda}}_l  \Tilde{\mathbf{Q}}_l^\top$ and the mixed product properties, we can write 
  \begin{equation}\label{eq: eigendecomposition saturated Kronecker}
      \mathbf{K}  = \left(\bigotimes_{l=1}^P \Tilde{\mathbf{Q}}_l \right)\left(\bigotimes_{l=1}^P \Tilde{\boldsymbol{\Lambda}}_l \right) \left(\bigotimes_{l=1}^P \Tilde{\mathbf{Q}}_l \right)^\top.
  \end{equation}
Note that $\Tilde{\mathbf{Q}} := \left( \bigotimes_{l=1}^P \Tilde{\mathbf{Q}}_l \right)$ is orthonormal and $\Tilde{\boldsymbol{\Lambda}} := \left(\bigotimes_{l=1}^P \Tilde{\boldsymbol{\Lambda}}_l \right)$ is diagonal with non-negative entries as each $\Tilde{\mathbf{K}}_l$ is positive semi-definite. Therefore, we can apply the Kronecker methods derived from \eqref{eq: eigendecomposition of gram matrix example}.

\subsubsection{The nonsaturated case}

We now show that the Kronecker product structure can be exploited for efficient computation even when we have a more general structure in the kernel, by using a empirically centred kernel \eqref{eq: empirically centred kernel}. Consider a hierarchical ANOVA kernel as in def. \ref{hierarchical_ANOVA_kernel} for data with a $P$-dimensional grid structure. Let us now assume that we have $|\mathcal{S}|$ terms in our additive kernel. The corresponding Gram matrix can be given by 
\begin{align}\label{eq: non-separble kronecker Gram matrix}
    \mathbf{K}=\sum_{u\in\mathcal{S}}\mathbf{K}_u, \quad
    \mathbf{K}_u = \bigotimes_{l=1}^P \mathbf{B}_l \ \text{ where } \notag\\ \mathbf{B}_l =
    \begin{cases}
        \mathbf{K}^{(c)}_{l},  &\text{if } l\in u \\
    \mathbf{1}_{n_l}\mathbf{1}_{n_l}^\top, &\text{otherwise}. 
    \end{cases}
\end{align}
For this class of Gram matrices, we prove the following theorem. 
\begin{theorem}\label{thm: eigen decomposition of additive centred kernel}
A matrix $\mathbf{K}$ of the form given by \eqref{eq: non-separble kronecker Gram matrix} has the following decomposition:
\begin{equation*}\label{eq: eigendecomposition }
    \mathbf{K} = \left(\bigotimes_{l=1}^P {\mathbf{Q}}_l^{(c)} \right)\mathbf{D}\left(\bigotimes_{l=1}^P {\mathbf{Q}}_l^{(c)} \right)^\top
\end{equation*}
where $\mathbf{Q}_l^{(c)}$ is orthonormal matrix whose columns consist of eigenvectors of $\mathbf{K}_l^{(c)}$, and $\mathbf{D}$ is diagonal with non-negative entries.
\end{theorem}

The proof of the theorem relies on the eigendecomposition of each $\mathbf{K}_l^{(c)}$, which is illustrated in the following proposition. Its proof can be found in Appendix \ref{apx: centred kernel eigen decomposition}.

\begin{proposition}\label{thm: eigen decomposition of a centred matrix}
Any $n\times n$ centred Gram matrix $\mathbf{K}^{(c)}$ has the following eigendecomposition:
    \begin{equation} \label{eq: eigen decomposition of centred kernel - apx}
         \mathbf{K}^{(c)} = \mathbf{Q}^{(c)}\boldsymbol{\Lambda}^{(c)}\mathbf{Q}^{(c)\top}\notag
    \end{equation}
where 
    \begin{equation}\label{eq: eigenvalues matrix of a centred matrix - apx}
        \boldsymbol{\Lambda}^{(c)} = \mathbf{diag}\left((0,\lambda_2,\ldots,\lambda_n)^\top\right)\notag
    \end{equation}
where $\lambda_j\geq 0, \forall j\in\{2,\dots,n\}$ and 
    \begin{equation} \label{eq: orthonormal matrix of a centred matrix - apx}
       \mathbf{Q}^{(c)} =
        \begin{bmatrix}
        \begin{matrix}
        \frac{1}{\sqrt{n}} \\
        \vdots \\
        \frac{1}{\sqrt{n}}
        \end{matrix}
        & \mathbf{q}_2 & \cdots & \mathbf{q}_n
        \end{bmatrix}.
    \end{equation}
\end{proposition}

We are now able to prove \ref{thm: eigen decomposition of additive centred kernel}.

\begin{proof}[Proof of Theorem 1]
    From \ref{thm: eigen decomposition of a centred matrix}, for $l=1,\ldots,P$, we have:
    \begin{eqnarray}
    \mathbf{K}_l^{(c)} &= \mathbf{Q}^{(c)}_l \boldsymbol{\Lambda}_l^{(c)} \mathbf{Q}^{(c)\top}_l \label{eq: eigendecomposition of a centred matrix for kronecker method}\\
    \mathbf{1}_{n_l}\mathbf{1}_{n_l}^\top &=  \mathbf{Q}^{(c)}_l \mathbf{A}_l\ \mathbf{Q}^{(c)\top}_l \nonumber
    \end{eqnarray}
    where 
    $\mathbf{Q}_l^{(c)}$ is orthonormal, $\boldsymbol{\Lambda}_l^{(c)}$ is diagonal with non-negative eigenvalues in the diagonal 
    and $\mathbf{A}_l$ is a $n_l\times n_l$ matrix with $\mathbf{A}_{1,1}=n_l$ and 0 everywhere else. Then using the mixed product property of Kronecker products, we can decompose $\mathbf{K}_u$ as 
    \begin{equation*}
        \mathbf{K}_u = \bigotimes_{l=1}^{P} \mathbf{Q}_{l}^{(c)}  \bigotimes_{l=1}^{P}\mathbf{D}_{ul}  \bigotimes_{l=1}^{P} \mathbf{Q}_{l}^{(c)\top}  
    \end{equation*}
    where 
    \begin{equation*}
        \mathbf{D}_{ul}= 
        \begin{cases}
            \boldsymbol{\Lambda}_l^{(c)} &\text{if} \  l\in u  \ \\
            \mathbf{A}_l &\text{otherwise}.
        \end{cases}
    \end{equation*}
    Let $\mathbf{D}_u = \bigotimes_{l=1}^{P} \mathbf{D}_{ul}$. We have
    \begin{align}
        \mathbf{K} &= \sum_{u\in\mathcal{S}} \mathbf{K}_u = \sum_{u\in\mathcal{S}}  \left( \bigotimes_{l=1}^{P} \mathbf{Q}_{l}^{(c)}  \mathbf{D}_u\bigotimes_{l=1}^{P} \mathbf{Q}_{l}^{(c)\top}\right)   \notag\\
        &= \bigotimes_{l=1}^{P} \mathbf{Q}_{l}^{(c)} \left(\sum_{u\in\mathcal{S}} \mathbf{D}_u\right) \bigotimes_{l=1}^{P} \mathbf{Q}_{l}^{(c)\top}.  \label{eq: eigendecomposition hierarchical Kronecker}
    \end{align}
    Since each $\mathbf{D}_{ul}$ is diagonal, also the matrix $\mathbf{D} := \sum_{u\in\mathcal{S}}  \mathbf{D}_u$ is diagonal with non-negative diagonal entries.
\end{proof}
\subsection{Computational complexity and space requirement} \label{sec: kronecker comp complexity}
Kronecker methods significantly reduce the cost of computing the log determinant of the matrix $\mathbf{K}+\sigma^2\mathbf{I}$, and solving the linear system $(\mathbf{K}+\sigma^2\mathbf{I})^{-1}\mathbf{v}$, which usually has $O(n^3)$ when $\mathbf{K}$ is an $n\times n$ Gram matrix. As seen in \eqref{eq: kronecker log determinant} and \eqref{eq: kronecker linear system}, the key operations are eigendecomposition to get eigenvalues and a matrix of eigenvectors, and matrix-vector multiplication involving Kronecker products. With a Kronecker product structure, eigendecomposition is applied to each $\mathbf{K}_l$ of size $n_l\times n_l$ individually, which has $O(n_l^3)$ complexity. The total cost for the eigendecomposition of $\mathbf{K}$ then reduces to $O(\sum_{l=1}^P n_l^3)$, which is dominated by the largest of $n_l$. The second component is a matrix-vector multiplication in $(\bigotimes_{l=1}^d\mathbf{Q}_l^\top)\mathbf{v}$. A matrix-vector multiplication of an $n\times n$ matrix and a vector of length $n$ usually requires $O(n^2)$ operations. Using the algorithm provided in \citet[Chapter 5]{saatcci2012scalable} and \citet{wilson2014fast}, this Kronecker product matrix-vector multiplication takes $O(n\sum_{l=1}^P n_l)$ which is much less than the usual $O(n^2)$. Once we have eigenvalues of all sub-Gram matrices $\mathbf{K}_l$, computing the log-determinant has an additional cost of $O(n)$. The storage requirement reduces from $O(n^2)$ to $O(\sum_{l=1}^P n_l^2)$ which is associated with storing matrices $\mathbf{Q}_1,\ldots,\mathbf{Q}_d$. 
Previous work by \citet{saatcci2012scalable} and \citet{wilson2014fast} explored the use of the Kronecker method in Gaussian process regression and demonstrated improved computational time through simulation studies. Our approach, which shares the same key factors determining computational cost (namely, eigendecomposition of Gram matrices and matrix-vector multiplication), is expected to yield similar computational gains.
\subsection{Other scalable approaches to GP models}\label{sec: other scalable approach}
A number of methods have been proposed to enhance the scalability of Gaussian process models. As summarized by \citet{liu2020gaussian}, one mainstream approach involves approximating the Gram matrix $\mathbf{K}$. This can be achieved by utilizing a subset of data, typically of size $m\ll n$ (subset-of-data), or by exploiting sparsity in the Gram matrix.  This is based on the assumption that the covariance between distant points is zero, resulting in sparse kernels \citep{melkumyan2009sparse}. A particularly popular technique is the low-rank approximation using inducing points  (e.g., \citet{titsias2009variational, hensman2013gaussian}), inspired by Nystrom's method \citep{williams2001using}. 
In spatio-temporal setting, \citet{datta2016nonseparable} introduced dynamic nearest neighbour GP that induces a sparse structure in the inverse of the covariance matrix with additive kernel structures.  Unlike the Kronecker approach, which necessitates a multi-dimensional grid structure for the data, these methods can be applied to broad data structure. However, the Kronecker approach offers the advantage of avoiding approximation, as it rather exploits the structure of the data to efficiently evaluate and store the key components required for estimation and inference. In fact, the Kronecker approach and other scalable methods can complement each other, as exemplified by \citet{wilson2015kernel}, who incorporated a grid structure into inducing points. Although their work focused on the tensor product kernel, the method can be extended to handle additive kernels using the decomposition discussed in Section \ref{sec: kronecker hierarchical}.

\section{Simulation studies}\label{sec: simulation}
We conduct simulation studies with two main aims.
First, we assess the ability of the proposed framework to identify the underlying interaction structure from a set of candidate additive-interaction GP models. In this setting, data are generated from a pre-specified additive hierarchical model, and multiple candidate models are fitted to evaluate model selection performance using several criteria, including the marginal likelihood, LOO log predictive density, and test mean absolute error (MAE).
Second, we examine the computational efficiency of the proposed Kronecker approach. Running the full model selection pipeline, which involves fitting multiple candidate models and computing metrics for model-fit or predictive performance, can be computationally demanding. Previous Kronecker-based GP approach was limited to either saturated models or those containing only the highest-order interaction term, which constrained their ability to explore and identify interaction structures in the data. We extend the use of Kronecker trick to more general hierarchical additive interaction models, which makes exploration of interaction structures for large-scale data computationally feasible.

\subsection{Simulation Setting}
We consider three predictors, $x = (x_1, x_2, x_3)$, each defined over the interval $[-5,5]$. For each dimension $\ell \in \{1,2,3\}$, a regular grid of points is first generated, and a subset of $n_\ell$ points is selected to construct the training grid, which forms a possibly irregular sub-grid in the three-dimensional space. The remaining grid points are used to randomly select test samples for model evaluation.
The response is generated as
\begin{equation}
y = f(x) + \epsilon, \qquad \epsilon \sim \mathcal{N}(0, \sigma^2), \notag
\end{equation}
where $f\sim \GP(0,k)$ and  
\begin{equation}
k(x, x') = \alpha_0^2 \sum_{u \in \mathcal{S}} k_u(x_u, x_u').\notag
\end{equation}
The base kernels are squared exponential,
\begin{equation}
k_j(x_j, x_j') = \alpha_j^2 \exp\!\left(-\frac{\|x_j - x_j'\|^2}{2\rho_j^2}\right),\notag
\end{equation}
with $\rho_j = 2.5$ and $\alpha_j = 0.7$ for $j = 1,2,3$. We set $k_{\emptyset}(x_\emptyset, x_\emptyset') = 1$ for the constant term. The parameters $\alpha_0$ and $\sigma$ vary depending on the simulation condition. The values of the parameters $\alpha_j$ are set such that the higher order interactions have smaller effects.

We consider five hierarchical models with increasing interaction complexity: 
(1) a main-effects-only model, 
(2) a model including one two-way interaction, 
(3) a model including two two-way interactions, 
(4) a model including all three two-way interactions, 
and (5) a saturated model including all main and interaction effects. 
The corresponding interaction structures $\mathcal{S}$ are summarised in Table~\ref{tab:models}.
\begin{table}[t]
\centering
\caption{Hierarchical models considered for the true data-generating process.}
\label{tab:models}
\begin{tabular}{ll}
\toprule
Model & Interaction structure $\mathcal{S}$ \\
\midrule
$\mathcal{M}_1$ & $\{\emptyset, \{1\}, \{2\}, \{3\}\}$ \\
$\mathcal{M}_2$ & $\{\emptyset, \{1\}, \{2\}, \{3\}, \{1,2\}\}$ \\
$\mathcal{M}_3$ & $\{\emptyset, \{1\}, \{2\}, \{3\}, \{1,2\}, \{2,3\}\}$ \\
$\mathcal{M}_4$ & $\{\emptyset, \{1\}, \{2\}, \{3\}, \{1,2\}, \{2,3\}, \{1,3\}\}$ \\
$\mathcal{M}_5$ & $\{\emptyset, \{1\}, \{2\}, \{3\}, \{1,2\}, \{2,3\}, \{1,3\}, \{1,2,3\}\}$ \\
\bottomrule
\end{tabular}
\end{table}
We compare the proposed hierarchical additive-interaction GP model fitted under the true generating structure against two nested neighbouring (simpler and more complex) ones. 
For instance, if the true data-generating model is $\mathcal{M}_4$, we fit $\mathcal{M}_3$ (simpler model), $\mathcal{M}_4$ (true model), and $\mathcal{M}_5$ (more complex model). 
Model performance is evaluated in terms of both fit and predictive accuracy. 

\paragraph{Simulation Study 1.}
This simulation study examines whether the proposed framework can correctly recover the underlying interaction structure from a set of candidate additive-interaction GP models. 
We set $n_\ell = 20$ for $\ell = 1,2,3$, giving a total of $N = 8{,}000$ training points, and randomly select $2{,}000$ test points for each replicate. 
Model fitting is performed using MCMC sampling implemented in \texttt{Stan} with the No-U-Turn sampler. 
Each model is estimated using four chains of 800 iterations, comprising 300 warm-up iterations and 500 post-warm-up samples. 
Model performance is assessed using best-fit and LOO log predictive density, DIC, log marginal likelihood computed using \texttt{bridgesampling} package and test MAE. 
Two noise levels are considered: $(\alpha_0, \sigma) = (1.0, 1.0)$ representing moderate noise and $(0.7, 1.0)$ representing relatively high noise. Each experimental scenario is repeated 500 times, and we report the proportion of correctly identified models.

\paragraph{Simulation Study 2.}
The second simulation study evaluates the computational efficiency of the proposed Kronecker-based inference compared with a naive (non-Kronecker) implementation. 
We follow a similar experimental setup to Simulation~1, using the same data-generating process as in the moderate-noise condition, but vary the total sample size $n$. 
For each true data-generating model, we record the total computation time required to fit three candidate models and compute selected model comparison metrics.

Due to computational cost, this experiment is run with 20 replicates, and the reported time is averaged over these runs. The variability across replicates, measured by the coefficient of variation (CV), remains below 20\% for most settings and below 30\% for all except the smallest Kronecker cases ($n \leq 1{,}000$), where absolute times are under one second and the relative variation is practically negligible (see Appendix \ref{apx : simulation}). 
We consider a single data-generating structure corresponding to $\mathcal{M}_3$, and fit three models: $\mathcal{M}_2$ (simpler), $\mathcal{M}_3$ (true), and $\mathcal{M}_4$ (more complex). 
Hyperparameters are estimated by maximising the log marginal likelihood in both implementations, as MCMC inference becomes infeasible for larger $N$ in the naive case. 
Since the model fitting is not done in a fully Bayesian manner and Simulation~1 showed that model-fit criteria outperform predictive metrics, we restrict comparison to the the best-fit and LOO log predictive density and DIC.
Sample sizes of $n = 125, 512,$ and $1{,}000$ are used for the naive approach, while the Kronecker approach is additionally evaluated for $n = 8{,}000$, $27{,}000$, $125{,}000$, and $1{,}000{,}000$ to demonstrate scalability.

\subsection{Results}
\paragraph{Simulation 1.}
\begin{table*}[!h]
\centering
\caption{The results of Simulation~1 showing the proportion of times (\%) the true model was correctly selected when compared against alternative candidate models (simpler and more complex) using different evaluation metrics. 
“Best-fit” denotes $\log p(\mathbf{y}|\bfx,\theta,\sigma^2)$ evaluated at the posterior mean of the hyperparameters, “LOO” refers to the leave-one-out log predictive density, and “Bridge” represents the log marginal likelihood computed using the \texttt{bridgesampling} package in \texttt{R}. 
Setting~1 corresponds to a medium-noise scenario $(\alpha_0 = 1.0, \sigma = 1.0)$, and Setting~2 corresponds to a relatively high-noise scenario $(\alpha_0 = 0.7, \sigma = 1.0)$.}
\resizebox{\textwidth}{!}{%
\begin{tabular}[t]{cccccccccccc}
\toprule
\multicolumn{2}{c}{ } & \multicolumn{2}{c}{Best-fit} & \multicolumn{2}{c}{DIC} & \multicolumn{2}{c}{LOO} & \multicolumn{2}{c}{Bridge} & \multicolumn{2}{c}{MAE} \\
\cmidrule(l{3pt}r{3pt}){3-4} \cmidrule(l{3pt}r{3pt}){5-6} \cmidrule(l{3pt}r{3pt}){7-8} \cmidrule(l{3pt}r{3pt}){9-10} \cmidrule(l{3pt}r{3pt}){11-12}
Setting & Model & simple & complex & simple & complex & simple & complex & simple & complex & simple & complex\\
\midrule
1 & $\mathcal{M}_1$ & NA & 96.7 & NA & 97.9 & NA & 92.8 & NA & 99.4 & NA & 71.3\\
  & $\mathcal{M}_2$ & 100.0 & 97.4 & 100.0 & 98.0 & 100.0 & 93.1 & 100.0 & 98.6 & 99.6 & 72.2\\
  & $\mathcal{M}_3$ & 100.0 & 98.6 & 100.0 & 98.6 & 100.0 & 94.6 & 100.0 & 99.4 & 99.4 & 72.2\\
  & $\mathcal{M}_4$ & 100.0 & 95.2 & 100.0 & 96.0 & 100.0 & 88.6 & 100.0 & 96.8 & 99.6 & 73.6\\
  & $\mathcal{M}_5$ & 100.0 & NA & 100.0 & NA & 100.0 & NA & 99.8 & NA & 96.4 & NA\\
\midrule
2 & $\mathcal{M}_1$ & NA & 90.7 & NA & 93.3 & NA & 88.0 & NA & 98.0 & NA & 68.0\\
  & $\mathcal{M}_2$ & 99.6 & 94.2 & 98.9 & 96.3 & 100.0 & 88.8 & 98.9 & 98.1 & 92.6 & 66.6\\
  & $\mathcal{M}_3$ & 100.0 & 95.6 & 99.8 & 96.2 & 99.8 & 89.3 & 99.8 & 97.8 & 95.5 & 68.0\\
  & $\mathcal{M}_4$ & 98.8 & 75.8 & 98.6 & 78.0 & 99.8 & 75.0 & 98.0 & 81.2 & 93.6 & 60.4\\
  & $\mathcal{M}_5$ & 77.2 & NA & 75.8 & NA & 77.0 & NA & 73.2 & NA & 68.0 & NA\\
\bottomrule
\end{tabular}%
}
\label{tab: sim1 result}
\end{table*}
Table~\ref{tab: sim1 result} summarises the results of Simulation~1, showing the proportion of times the true model was correctly selected under two noise settings. 
Setting~1 corresponds to the medium-noise scenario, while Setting~2 represents the relatively high-noise scenario. 
Across both settings, candidates model are compared based on several model fit and predictive metrics: the $\log p(\mathbf{y}|\bfx,\theta,\sigma^2)$ evaluated at the posterior mean of the hyperparameters (“Best-fit”), DIC, the LOO log predictive density, the log marginal likelihood estimated via bridge sampling (“Bridge”), and the test MAE.

In Setting~1, selecting the correct model against a simpler alternative is almost always successful, with proportions close to 100\% across all metrics. 
Distinguishing the true model from a more complex alternative, i.e., one that includes additional interaction terms, appears slightly more challenging, but accuracy remains high, typically above 95\%, with the lowest observed proportion being 88.6\% for the LOO metric. 
In contrast, the test MAE yields consistently lower selection rates, particularly when comparing the true model against more complex alternatives, suggesting that the predictive accuracy alone may lead to selecting an necessarily larger model.

Under the relatively high-noise condition (Setting~2), the overall proportion of correct selections remains high but declines for models involving higher-order interaction terms. 
Identifying the true model containing all two-way interactions against the saturated model including a three-way term—or vice versa—becomes more difficult. 
Nevertheless, model fit metrics (Best-fit, DIC, LOO, and Bridge) retain substantially higher selection rates compared with test MAE, which again shows reduced sensitivity under noisier conditions. 
These results suggest that likelihood-based criteria remain more robust to noise and provide more reliable guidance for distinguishing between hierarchical interaction structures of varying complexity.

\begin{figure}[t]
    \centering
    \includegraphics[width=1.0\linewidth]{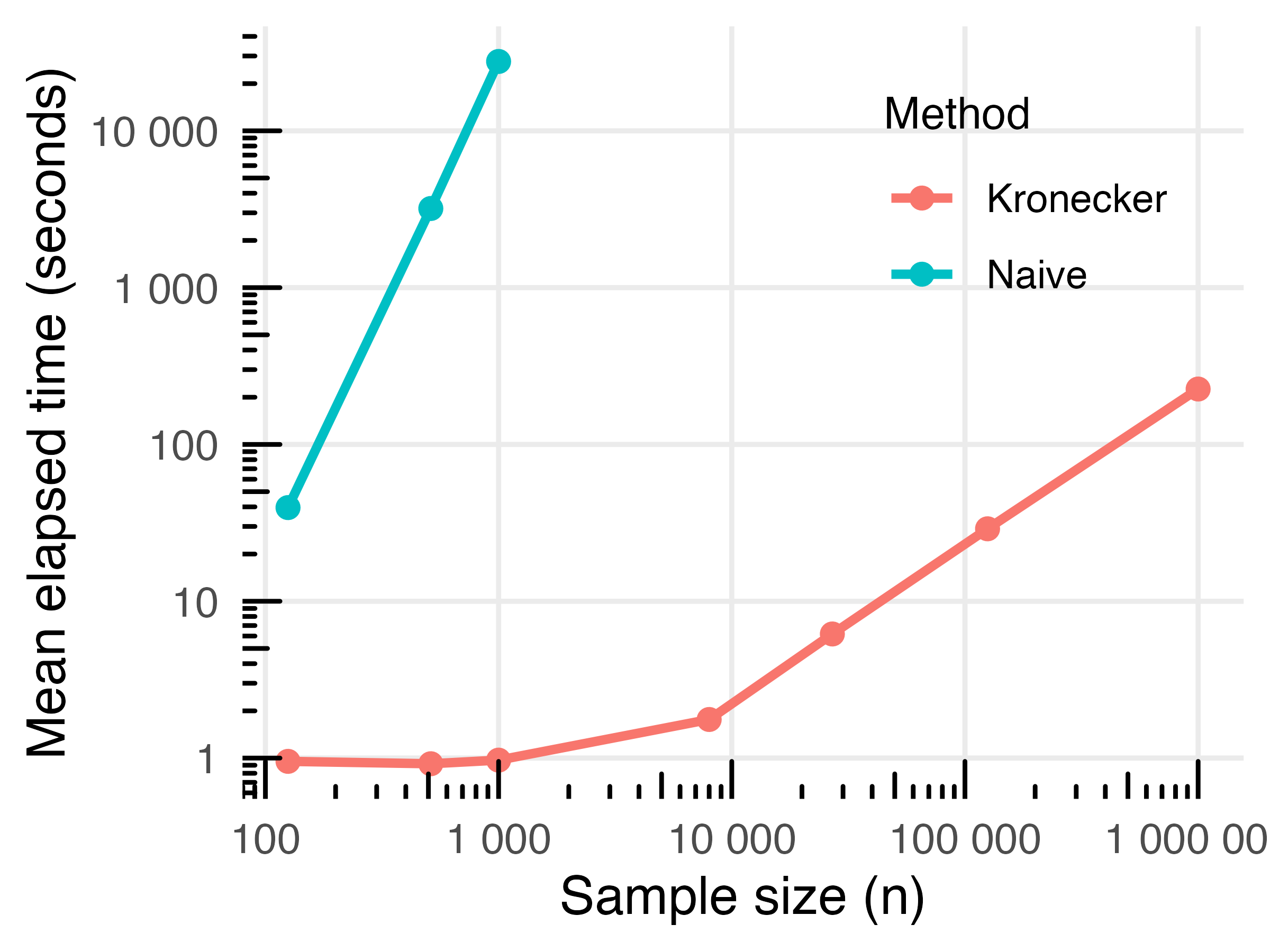}
    \caption{The results of Simulation~2 showing computation time (in seconds) for fitting three candidate models and computing the LOO and best-fit log likelihood, averaged over 20 replicates. 
    The proposed Kronecker implementation and the naive (non-Kronecker) implementation are compared across different sample sizes $n$ on a log--log scale. }
    \label{fig:time_scaling}
\end{figure}

\paragraph{Simulation 2.}
Figure~\ref{fig:time_scaling} presents the results of the second simulation study, comparing computation time between the proposed Kronecker implementation and the naive (non-Kronecker) approach. 
The time reported corresponds to fitting three candidate models and computing the LOO and best-fit log predictive density, averaged over 20 replicates.

The results clearly demonstrate the substantial computational advantage of the Kronecker approach. 
While the naive implementation becomes infeasible beyond $n = 1{,}000$, taking on average more than seven hours for this case, the Kronecker implementation handles substantially larger datasets with ease. 
Even at $n = 1{,}000{,}000$, the Kronecker method completes in under four minutes on average, exhibiting near-linear scaling with respect to $n$ in practical ranges. 
These results indicate that exploiting the Kronecker structure substantially improves computational efficiency, allowing model fitting and comparison to remain feasible even for large-scale datasets. 

\section{Real-world application}\label{sec: application}
To illustrate the benefit of the proposed method, we analyse hourly nitrogen dioxide ($\text{NO}_2$) concentration data from the London Air Quality Network, covering the period from January to May 2020, which includes the onset of the first COVID-19 lockdown in the UK. $\text{NO}_2$ is a major air pollutant associated with vehicle emissions and is known to adversely affect respiratory and cardiovascular health, as well as contribute to urban smog and acid deposition. Understanding short-term variation and spatial patterns in $\text{NO}_2$ is important for evaluating the effects of emission control policies and behavioural changes such as reduced mobility during lockdown periods.
A number of studies have shown that the COVID-19 lockdowns contributed to a notable drop in $\text{NO}_2$ concentrations worldwide \citep[e.g.][]{dutta2021recent,cooper2022global} and in the UK specifically \citep{lee2020uk,jephcote2021changes}. Most analyses have been based on daily or weekly mean data, which are suitable for studying long-term trends. In contrast, hourly data provide richer information on diurnal variation and allow us to examine how daily cycles evolved during this period. Analysing such high-frequency, spatially distributed data requires methods that can efficiently capture spatio-temporal dependencies and scale to large structured datasets.

The dataset used here contains hourly $\text{NO}_2$ measurements (in $\mu g/m^3$) from 59 monitoring sites across Greater London, after excluding stations with excessive missingness. The data form a balanced three-dimensional structure indexed by site, day, and hour of day, resulting in approximately 208,000 observations. Missing values (less than one percent) were imputed to form a complete grid suitable for Kronecker-based computation. The measurement sites are classified into five categories: Kerbside, Roadside, Urban Background, Suburban, and Industrial. Further details of the dataset and preprocessing steps are provided in Appendix~\ref{apx: dataset issues}.

Although the analysis is exploratory, it highlights the scalability and interpretability of the proposed approach when applied to large structured spatio-temporal data.

\subsection{Model formulation}\label{sec: model formulation}
The dataset has a three-dimensional grid structure, with location, day and hour as predictors, denoted by $\mathbf{x}_1 \in \mathcal{X}_1$, $x_2\in\mathcal{X}_2$ and $x_3\in \mathcal{X}_3$ where $\mathcal{X}_1\subset\mathbb{R}^2$, $\mathcal{X}_2$ represents the set of calendar dates numbered $1,2,\ldots$, and $\mathcal{X}_3$ represent hour of the day indexed by ${1,2,...,24}$. We write $\mathcal{X} = \mathcal{X}_1 \times \mathcal{X}_2 \times \mathcal{X}_3$.
The response variable $y_{s,d,h}$ denotes the observed $\text{NO}_2$ concentration at site $s$, on day $d$, and hour $h$, where $s=1,\ldots,n_1$, $d=1,\ldots,n_2$, and $h=1,\ldots,n_3$ with $n_1=59$, $n_2=147$, and $n_3=24$. We model
\begin{equation*}
    y_{s,d,h} = f(\mathbf{x}_{1s},x_{2d},x_{3h}) + \epsilon_{s,d,h},
    \quad \epsilon_{s,d,h} \sim \mathcal{N}(0,\sigma^2),
\end{equation*}
where $f \sim \GP(0,k)$ with the structure of the kernel $k:\mathcal{X}\times\mathcal{X}\to\mathbb{R}$ determined by the choice of the model. We fitted several hierarchical additive interaction GP models, differing by which main and interaction effects are included. Specifically, we considered (1) a main-effect model, (2) a model with spatial interactions with day and hour, (3) a model including all two-way interactions, (4) a saturated model with a three-way interaction term, and (5) a separable model with the three-way interaction only. The corresponding kernel constructions are summarised in Table~\ref{tab:model-summary-no2}.
\begin{table}[t]
\centering
\setlength{\tabcolsep}{1.5pt}
\caption{Summary of Gaussian process models fitted to the \(\text{NO}_2\) dataset.
Each model assumes \( f \sim \mathcal{GP}(0, \alpha_0^2 k_m) \), where \(k_m\) denotes the kernel structure below.
Base kernels \(k_l\) (\(l=1,2,3\)) correspond to spatial location, day, and hour, respectively.}
\begin{tabular}{ll}
\toprule
\textbf{Model} & \textbf{Kernel structure} \\
\midrule
1: Main effects only & $k_{m1} = 1 + k_{1} + k_{2} + k_{3}$ \\[3pt]
2: Space--time interactions & $k_{m2} = k_{m1} + k_{1}\otimes (k_{2} + k_{3})$ \\[3pt]
3: All two-way interactions & $k_{m3} = k_{m2} + k_{2}\otimes k_{3}$ \\[3pt]
4: Saturated model & $k_{m4} = k_{m3} + k_{1}\otimes k_{2}\otimes k_{3}$ \\[3pt]
5: Three-way interaction only & $k_{m5} = k_{1}\otimes k_{2}\otimes k_{3}$ \\
\bottomrule
\end{tabular}
\label{tab:model-summary-no2}
\end{table}

Each base kernel $k_l$ ($l=1,2,3$) corresponds to one predictor and was defined using a squared and centred fractional Brownian motion kernel (see Appendix~\ref{apx: kernel choice} for details). Hyperpriors for the scale and noise parameters were specified as described in the same appendix. All models were implemented and estimated using \texttt{Stan}.

\subsection{Results}

\begin{table*}[t]
\centering
\caption{Results from fitting Model 1 to Model 5 to London $\text{NO}_2$ data. The difference of the log marginal likelihood in comparison to that of the baseline model (Model 1) is shown as $\Delta$mloglik. The log marginal likelihood for Model 1 is -858,044. The average time (in minutes) taken to obtain 4 chains of 1000 MCMC samples after the 500 warm-up phase is also displayed. For model 5, we only need one scale parameter $\alpha_1$ due to identifiability issues. }
\label{tab:NO2 results}
\begin{tabular}{lcccccrr}
\hline
Model & $\alpha_0$ & $\alpha_1$ & $\alpha_2$ & $\alpha_3$ & $\sigma$ &$\Delta$mloglik & Time  \\ \hline
1: main & 6.80 & 5.11  & 2.24 & 0.34 & 14.9 & - & 29.0  \\
2: spatio-temporal interaction& 14.4 & 1.28 & 0.67 & 0.24 & 12.5 & 25,794 & 41.9 \\
3: all two-way interaction & 10.6& 1.86& 1.32 & 0.31 & 8.37 & 98,118 & 59.1 \\
4: saturated & 52.0 & 0.48 & 0.94& 0.051 & 6.51 & 102,867 & 57.7 \\
5: three-way interaction only & - & 0.00081 & - & - & 36.4 & -186,450 & 1.96 \\
\hline
\end{tabular}
\end{table*}

The main results are summarised in Table~\ref{tab:NO2 results}, which reports the posterior mean of the hyperparameters, improvements in log marginal likelihood relative to the baseline main effect model, and computational time. The log marginal likelihood increases from Model~1 to Model~4, indicating progressively better model fit as interaction terms are added. Model~3 (all two-way interactions) shows the largest gain compared to Model~2, which reflects the importance of capturing the interaction between daily cycles and the global temporal pattern. The saturated Model~4 achieves the highest log marginal likelihood but with a moderate improvement over Model~3, while the three-way–interaction-only Model~5 performs considerably worse. All models were estimated within approximately 20~minutes, demonstrating that the proposed Kronecker-based inference enables efficient comparison across models of varying complexity. Additional results, such as posterior predictive plots for selected sites are provided in Appendix~\ref{apx:posterior-predictive}.

\begin{figure}[t]
    \centering
    \includegraphics[width=1\linewidth]{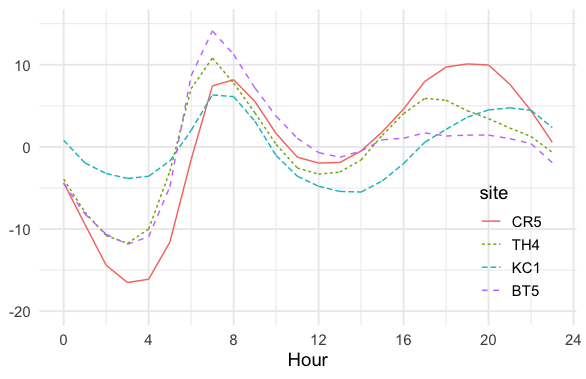}
    \caption{Hour-of-day effect on $\text{NO}_2$ concentration (in $\mu$g/m$^3$) at selected monitoring sites. Kerbside (CR5) and roadside (TH4) sites show pronounced morning and evening peaks, whereas the urban background site (KC1) exhibits smaller amplitudes and the industrial site (BT5) shows a single morning peak}
    \label{fig: no2 hour effect by site}
\end{figure}

\begin{figure*}[t]
\centering
  \begin{subfigure}[t]{.45\linewidth}
    \centering
    \includegraphics[width=\linewidth]{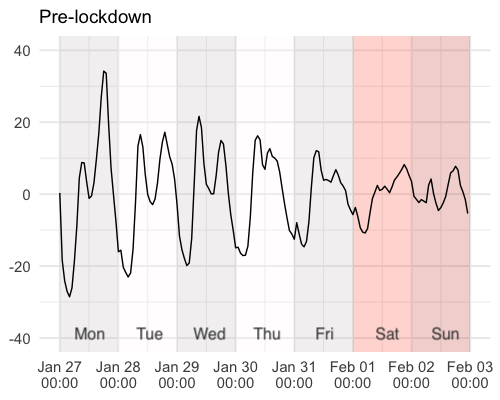}
  \end{subfigure}
  \hfill
  \begin{subfigure}[t]{.45\linewidth}
    \centering
    \includegraphics[width=\linewidth]{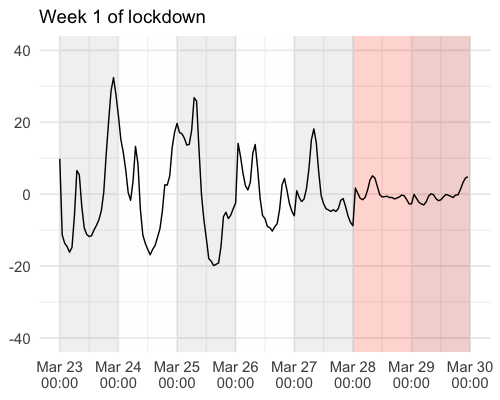}
  \end{subfigure}
  \begin{subfigure}[t]{.45\linewidth}
    \centering
    \includegraphics[width=\linewidth]{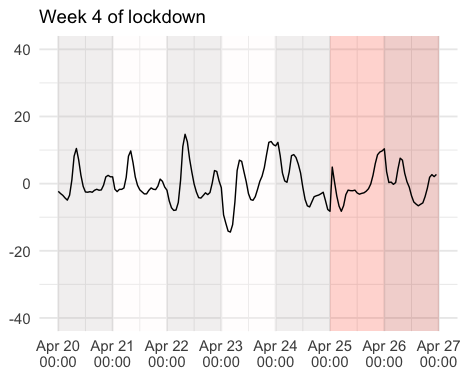}
  \end{subfigure}
  \hfill
  \begin{subfigure}[t]{.45\linewidth}
    \centering
    \includegraphics[width=\linewidth]{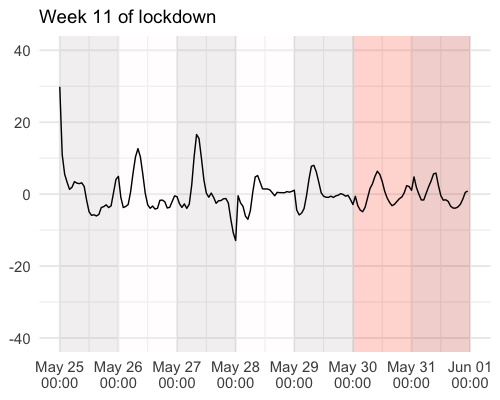}
  \end{subfigure}
\caption{Change in the hour-of-day effect (in $\mu$g/m$^3$) over time, averaged over monitoring sites. The evening peak gradually weakens after the onset of the COVID-19 lockdown.}
  \label{fig: interaction effect between day and hour}
\end{figure*}

The saturated model, which includes up to three-way interactions, was selected as it provided the best overall fit. Here, we focus on interpreting a subset of its effects related to diurnal variation. Figures~\ref{fig: no2 hour effect by site} and~\ref{fig: interaction effect between day and hour}, which respectively show the posterior means $\bar{m}_3(x_3) + \bar{m}_{13}(\bfx_1, x_3)$ and $\bar{m}_3(x_3) + \bar{m}_{23}(x_2, x_3)$ as functions of hour $x_3$ for different values of $\bfx_1$ and $x_2$, illustrate how the amplitude and shape of daily cycles vary across monitoring sites and evolve over the study period.
Due to the corresponding kernel functions in the prior, the posterior mean of the three-way interaction $\bar{m}_{123}(\bfx_1, x_2, x_3)$ satisfies a sum-to-zero property over each input dimension, e.g. $\sum_{d}\bar{m}_{123}(\bfx_1, x_{2d}, x_3) = 0$ and $\sum_{s}\bar{m}_{123}(\bfx_{1s}, x_2, x_3) = 0$. Although the selected model indicates that the effect of hour varies by site and that these site-specific diurnal patterns change over time, this property allows $\bar{m}_3(x_3) + \bar{m}_{13}(\bfx_1, x_3)$ to be interpreted as the site-specific diurnal pattern averaged over time. Analogously, $\bar{m}_3(x_3)+\bar{m}_{23}(x_2,x_3)$ can be interpreted as representing temporal changes in the diurnal pattern averaged over sites. 

\section{Conclusion}\label{sec: conclusion}
This paper presented a framework for hierarchical additive interaction modelling with a GP prior. Specifically, we used the hierarchical ANOVA decomposition kernel to represent main and interaction effects in a manner consistent with the principle of marginality. Kernel centring was incorporated to ensure identifiability and to provide interpretable estimates of main and interaction effects. When the data form a multi-dimensional grid, the resulting covariance matrix often exhibits a Kronecker product structure, enabling substantial computational gains. However, previous Kronecker-based methods were restricted to models that can be expressed as a single tensor product, such as interaction-only or fully saturated formulations. By exploiting kernel centring within the hierarchical ANOVA decomposition, we show that the Kronecker approach can be extended to handle any interaction structure under the proposed class of hierarchical additive GP models. This allows efficient, exact inference and model comparison for large multi-dimensional grid data.

The simulation studies demonstrated the computational performance and practical utility of the proposed methodology. Using three covariates, we confirmed the theoretical linear scaling of computation with respect to data size and showed that candidate models with different interaction structures can be fitted and evaluated efficiently. The results also indicated that test predictive accuracy alone may not reliably identify the correct model, particularly in high-noise settings. In contrast, likelihood-based criteria such as the marginal likelihood and leave-one-out predictive density provided more robust guidance for comparing models and identifying the true interaction structure.

In the real-world application to hourly nitrogen dioxide ($\text{NO}_2$) concentration data in London, the proposed approach enabled full Bayesian inference for all candidate models at a manageable computational cost. The identified main and interaction effects were interpretable and could be visualised to understand spatial, temporal, and periodic patterns of $\text{NO}_2$ concentration levels.

Several limitations remain. The computational efficiency of the proposed method relies on data conforming to a grid structure, and for large-scale datasets with more general data structures, other scalable GP approaches are required. Investigating model selection and evaluation procedures for such settings, where the Kronecker structure cannot be directly exploited, is an important direction for future work. Another avenue for further research concerns model selection strategy. In many applied contexts, model specification should be guided by domain knowledge and the underlying research question rather than by exhaustive model search, which was the approach adopted in our real-world application. Nevertheless, when the number of covariates is large, the space of possible models expands rapidly, and developing practical strategies for efficient model selection will be necessary.

Finally, we assumed complete data on a grid, and missing observations were handled through simple imputation. Approaches such as those by \citet{gilboa2013scaling} and \citet{wilson2014fast}, which approximate the likelihood in the presence of missing values, could be readily incorporated within our framework as approximations to the complete-data analysis. Future work should also consider more complex cases such as non-random or censored missingness, which remain challenging for current Gaussian process models.

\backmatter

\begin{appendices}

\section{Methodology}
\subsection{Centring of kernel}\label{apx: kernel constraints}
The centring operation shifts the kernel so that its mean is null. Indeed, for all $\bfx'\in\mathcal{X}$: 
\begin{align}
    \mu_{\tilde{k}}(\bfx')=&\E_{X\sim\nu}\left[\tilde{k}(X,\bfx')\right]\notag\\
    =& \E_{X\sim\nu}\left[k(X,\bfx')-\E_{X'\sim\nu}\left[k(\bfx,X')\right]\right.\notag\\
    &\left.-\E_{X\sim\nu}\left[k(X,\bfx')\right]+\E_{X,X'\sim\nu}\left[k(X,X')\right]\right]\notag\\
    =&\E_{X\sim\nu}\left[k(X,\bfx')\right]-\E_{X\sim\nu}\left[k(X,\bfx')\right]\notag\\
    =&0
\end{align}
Then, every function living in the RKHS generated by $\tilde{k}$ satisfies the following:
\begin{align}
    \E_{X\sim\nu}\left[f(X)\right]&=\int f(\bfx)\nu(d\bfx)=\int \langle f, \tilde{k}(\bfx,\cdot)\rangle\nu(d\bfx)\notag\\
    &=\langle \int \tilde{k}(\bfx,\cdot)\nu(d\bfx), f\rangle= \langle \mu_{\tilde{k}}(\bfx), f \rangle \notag\\
    &= \langle 0, f \rangle = 0.
\end{align}

\subsection{Eigendecomposition of a centred Gram matrix} \label{apx: centred kernel eigen decomposition} 
Consider a $n\times n$ Gram $\mathbf{K}^{c}$ matrix given by an emprically centred kernel function \eqref{eq: empirically centred kernel}. In what follows we prove that a centred Gram matrix has an eigendecomposition of a special form, as stated in Proposition \ref{thm: eigen decomposition of a centred matrix}. We write an eigendecomposition of a matrix $\mathbf{M}$ by $\mathbf{M} = \mathbf{Q}\boldsymbol{\Lambda}\mathbf{Q}^\top $ where $\boldsymbol{\Lambda}$ is a diagonal matrix of which diagonal element are eigenvalues of $\mathbf{M}$ in non-decreasing order and $\mathbf{Q}$ is an orthonormal matrix with its $i$-th column $\mathbf{q}_i$ being the eigenvector which corresponds to $i$-th eigenvalue. First, we state and prove this result on eigenvectors of $\mathbf{K}^{c}$.

\begin{proposition}\label{thm: eigenvectors associated with non-zero eigenvalues of centred matrix}
Any eigenvector $\mathbf{q}_i$ of a $n\times n$ centred Gram matrix $\mathbf{K}^{(c)}$ associated with non-zero eigenvalue $\lambda_i$ is orthogonal to $\mathbf{1}_n$. 
\end{proposition}
\begin{proof}Using $\mathbf{K}^{(c)}\mathbf{q}_i=\lambda_i\mathbf{q}_i$, we have 
    \begin{equation*}
        \mathbf{q}_i^\top \mathbf{1}_n = \frac{1}{\lambda_i}\mathbf{q}_i^\top\mathbf{K}^{(c)}\mathbf{1}_n = \mathbf{0}.
    \end{equation*}
The last equality is due to the fact that all rows and columns of a centred matrix sum to 0. 
\end{proof}

Using the previous result, we can now prove Proposition \ref{thm: eigen decomposition of a centred matrix}.

\begin{proof}[Proof of Proposition \ref{thm: eigen decomposition of a centred matrix}.] 
Let $k$ denote the number of zero eigenvalues of $\mathbf{K}^{(c)}$. Due to the centring, $\text{rank}(\mathbf{K}^{(c)}) \leq n-1$, i.e., we have $k\geq1$.\\
For $k = 1$, we have $\lambda_j>0$, $\forall j \in \{2\ldots,n\}$ and the eigenvectors $\mathbf{q}_2,\ldots,\mathbf{q}_n$ are orthogonal to $\mathbf{1}_n$ from proposition \ref{thm: eigenvectors associated with non-zero eigenvalues of centred matrix}. Normalising the vector $\mathbf{1}_n$ completes an orthonormal basis, hence the first column of $\mathbf{Q}^{(c)}$ is given by $\frac{1}{\sqrt{n}}\mathbf{1}$. \\
For $k\geq2$, the first $k$ columns of $\mathbf{Q}^{(c)}$, $(\mathbf{q}_1,\ldots,\mathbf{q}_k)$, are not uniquely determined. Using $\mathbf{q}_j^\top(\frac{1}{\sqrt{n}}\mathbf{1}_n) = \frac{1}{\sqrt{n}}\mathbf{q}_j^\top\mathbf{1}_n = \mathbf{0}$ for $j=k+1,\ldots,n$, we set $\mathbf{q}_1 = \frac{1}{\sqrt{n}}\mathbf{1}_n$ and find $(\mathbf{q}_2,\ldots, \mathbf{q}_k)$ to complete an orthonormal system.
\end{proof}
In practice, we may use a computer program to obtain a initial set of normalised eigen-vectors denoted by $\mathbf{v}_1,\ldots,\mathbf{v}_n$. For $k\geq2$, $\mathbf{v}_1,\ldots,\mathbf{v}_k$ may not contain a vector $\frac{1}{\sqrt{n}}\mathbf{1}_n$ but $\text{span}(\mathbf{v}_1,\ldots,\mathbf{v}_k)$ contains $\mathbf{1}_n$. To have orthonormal bases $\mathbf{q}_1,\ldots,\mathbf{q}_n$ specified above, we take $\mathbf{q}_1 = \frac{1}{\sqrt{n}}\mathbf{1}_n$ and $\mathbf{q}_j = \mathbf{v}_j$ for $j = k+1,\ldots,n$. The rest of the vectors $\mathbf{q}_2,\ldots,\mathbf{q}_k$ can be computed using for example Gram–Schmidt process.
\begin{remark}
The $n\times n$ matrix $\mathbf{1}_{n}\mathbf{1}_n^\top$ has the following decomposition:
    \begin{equation}
        \mathbf{1}_{n}\mathbf{1}_n^\top = \mathbf{Q}^{(c)}\mathbf{A}_n\mathbf{Q}^{(c)\top} \label{eq: eigen decomposition of 1 matrix - apx}
    \end{equation}
where $\mathbf{Q}^{(c)}$ is given by (\ref{eq: orthonormal matrix of a centred matrix - apx}) and $\mathbf{A}_n$ is a $n\times n$ matrix with its $i,j$-th element equals $n$ for $i=j=1$ and 0 everywhere else, i.e., 
\begin{equation} \label{eq: eigen decomposition of 1 matrix, diagonal matrix - apx}
    \mathbf{A}_n = \begin{bmatrix}
        n & 0 & \cdots & 0 \\
        0 & 0 & \cdots & 0 \\
        \vdots & \vdots & \ddots & \vdots \\
        0 & 0 & \cdots & 0
        \end{bmatrix}.
\end{equation}
\end{remark}
\section{Simulation}\label{apx : simulation}
This section presents additional results from the simulation studies. 
Table~\ref{tab:timing_summary} summarises the results of Simulation~2, reporting average computation time (in seconds) and the coefficient of variation (CV) over 20 replicates. 
The timings correspond to the elapsed wall-clock time required to fit three candidate models and compute the LOO and best-fit log predictive density under both the Kronecker and naive implementations. 
Computation time increases with sample size $N$, but the Kronecker approach remains consistently faster across all settings, with moderate variability across replicates.

\begin{table}[t]
\centering
\caption{Average computation time (in seconds) and coefficient of variation (CV) over 20 replicates for each method and sample size $N$. 
The time show is elapsed wall-clock time required to fit three candidate models and compute LOO and best-fit log perdictive density.}
\label{tab:timing_summary}
\begin{tabular}{lrrr}
\toprule
Method & $N$ & Average time (s) & CV \\
\midrule
Kronecker &     125   &     0.953  & 0.352 \\
          &     512   &     0.920  & 0.303 \\
          &    1{,}000   &     0.970  & 0.388 \\
          &    8{,}000   &     1.762  & 0.159 \\
          &   27{,}000   &     6.185  & 0.193 \\
          &  125{,}000   &    28.978  & 0.226 \\
          &1{,}000{,}000 &   225.613  & 0.293 \\
\midrule
Naive     &     125   &    39.569  & 0.132 \\
          &     512   &  3{,}196.460  & 0.132 \\
          &   1{,}000   & 27{,}686.845 & 0.184 \\
\bottomrule
\end{tabular}
\end{table}

\section{Data analysis}
\subsection{Dataset}\label{apx: dataset issues}
\begin{figure}[]
    \centering
    \includegraphics[width=0.99\linewidth]{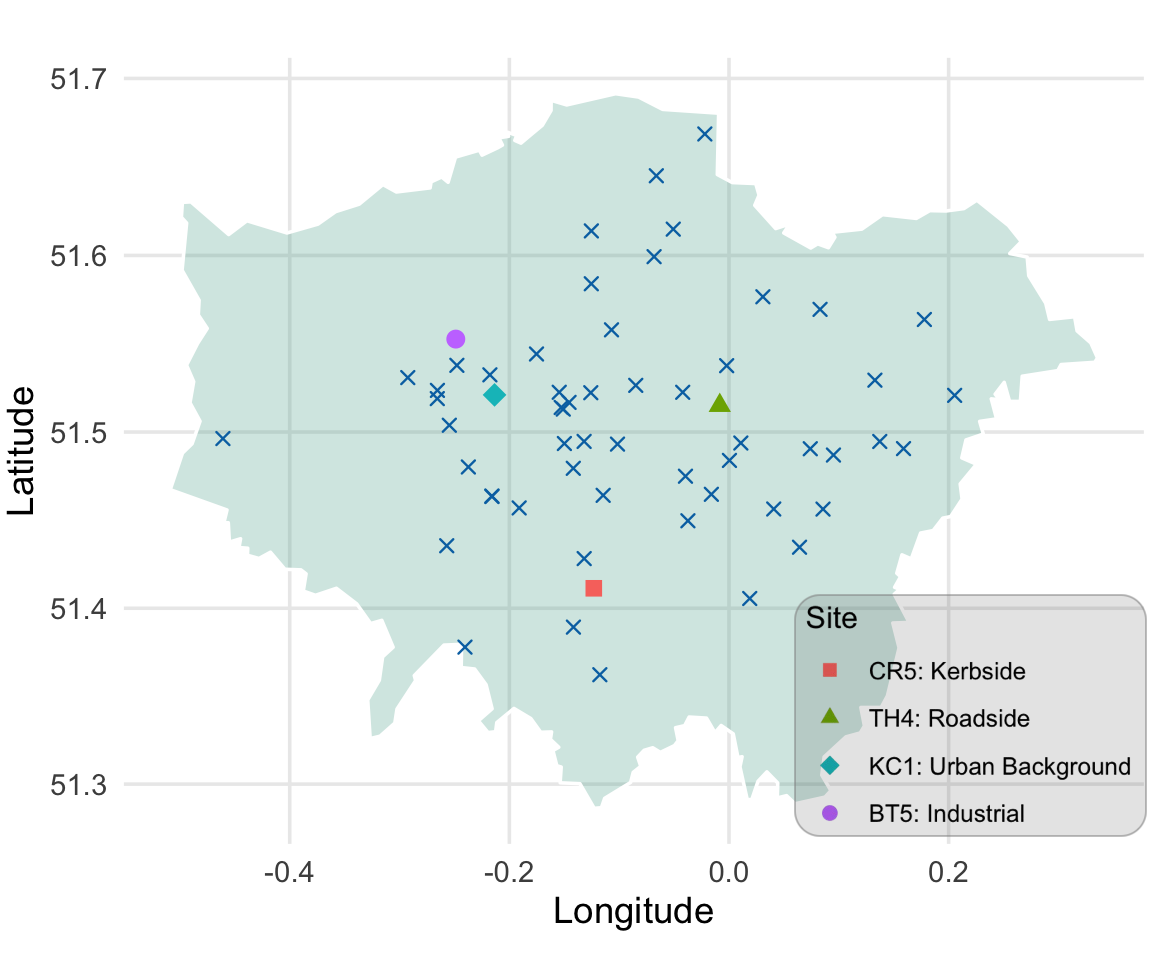}
    \caption{The location of $\text{NO}_2$ measurement sites included in the dataset. Four sites are selected and labelled for illustration purposes. Site CR5 is in the Borough of Croydon, while TH4, KC1, and BT5 are in Tower Hamlets, Kensington and Chelsea, and Brent, respectively. The sites are classified into five categories: Kerbside, Roadside, Urban Background, Suburban, and Industrial.}
    \label{fig: no2 site location}
\end{figure}

Figure~\ref{fig: no2 site location} shows the locations of the 59 monitoring sites included in the analysis, which are part of the London Air Quality Network\footnote{\url{https://www.londonair.org.uk}}. The sites span the Greater London area and are classified into five categories: Kerbside, Roadside, Urban Background, Suburban, and Industrial. Four representative sites are highlighted for illustration. 

The dataset comprises hourly $\text{NO}_2$ concentration measurements (in $\mu g/m^3$) recorded between January 6 and May 30, 2020, yielding 208,152 observations after excluding sites with more than 30\% missing data or with gaps exceeding 48 consecutive hours. The data exhibit a three-dimensional structure indexed by site, day, and hour of the day, forming a balanced spatio-temporal grid. A small proportion of observations (1,290 missing values out of 208,152, 0.62\%) were missing and were imputed to obtain a complete structure suitable for Kronecker-based computation. A simple approach was adopted due to the very low proportion of missing values. 

For each missing value at a measurement site, we created a small subset of the data consisting of the observations collected from the same site from 24 hours before to 24 hours after the missing value. A simple one-dimensional Gaussian process regression with a squared and centred standard ($\gamma=\frac{1}{2}$) Brownian Motion kernel was then fitted, and the missing value was replaced with the posterior predictive mean.

The study period, which spans from January 6 to May 31, 2020, includes the transition to British Summer Time (BST) beginning at 1:00 AM on March 29. The timestamps in the original data were all recorded in Greenwich Mean Time (GMT). We converted the timestamps to match BST from 1:00 AM (GMT) onwards, resulting in a one-hour gap at 1:00 AM in the adjusted time series. The gap was filled using the mean of the records immediately before and after. 
\subsection{Kernel choice}\label{apx: kernel choice}
For the analysis, we used the \textit{squared centred fractional Brownian motion} (fBM) kernel, with empirical centring applied to ensure identifiability of additive components.

The fractional Brownian motion kernel is defined for $\mathbf{x}, \mathbf{x}' \in \mathcal{X} \subset \mathbb{R}^p$, scale parameter $\alpha > 0$, and Hurst coefficient $0 < \gamma < 1$ as
\begin{equation}\label{eq:fbm-kernel}
k_{\text{fBM}_\gamma}(\mathbf{x}, \mathbf{x}') = \frac{\alpha^2}{2}\bigl(|\mathbf{x}|^{2\gamma} + |\mathbf{x}'|^{2\gamma} - |\mathbf{x} - \mathbf{x}'|^{2\gamma}\bigr).
\end{equation}
The Hurst coefficient $\gamma$ controls the roughness of sample paths: smaller values produce rougher processes, while larger values lead to smoother trajectories. The case $\gamma = 0.5$ corresponds to the standard Brownian motion kernel.

Let $k^{(c)}$ denote fBM kernel smpirically centred using \eqref{eq: empirically centred kernel}. To improve smoothness, we use the squared kernel construction
\begin{equation}\label{eq:squared-fbm}
k_{\text{sq}}(\mathbf{x}, \mathbf{x}') = \sum_{i=1}^n k^{(c)}(\mathbf{x}, \mathbf{x}_i) , k^{(c)}(\mathbf{x}', \mathbf{x}_i),
\end{equation}
yielding a {squared centred fBM kernel}. The resulting kernel remains positive definite and produces much smoother sample paths than the original fBM kernel. See \citet{bergsma2020regression} for the detail of the smoothness properties of fBM paths and squared fBM path.

In this study, we used $\gamma = 0.3$ for the spatial process $f_1$ and $\gamma = 0.5$ (the standard Brownian motion case) for the temporal processes $f_2$ and $f_3$. 

\subsection{Additional results}
\subsubsection{Posterior predictive}\label{apx:posterior-predictive}
\begin{figure*}[t]
  \begin{subfigure}[t]{.49\linewidth}
    \centering
    \includegraphics[width=\linewidth]{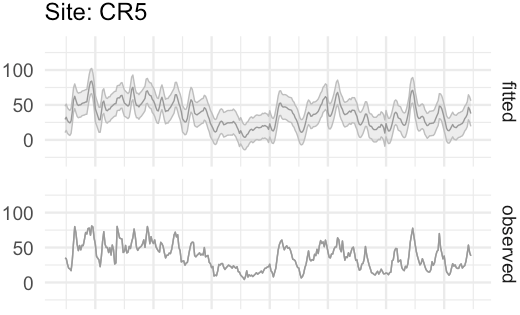}
  \end{subfigure}
  \hfill
  \begin{subfigure}[t]{.49\linewidth}
    \centering
    \includegraphics[width=\linewidth]{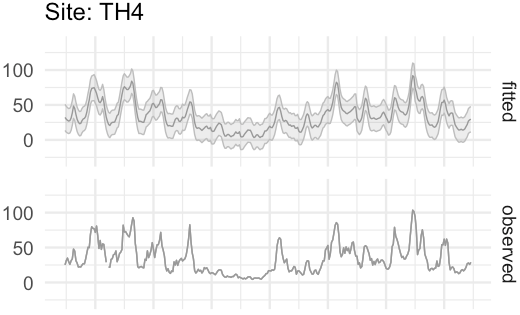}
  \end{subfigure}
  \medskip
  \begin{subfigure}[t]{.49\linewidth}
    \centering
    \includegraphics[width=\linewidth]{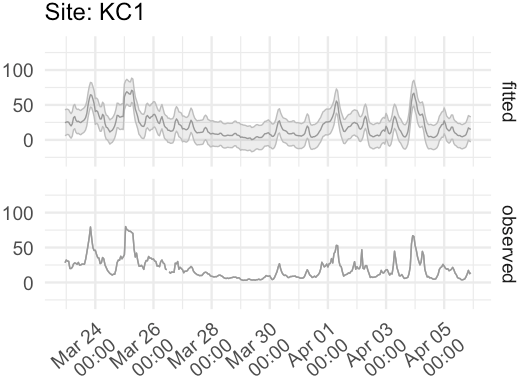}
  \end{subfigure}
  \hfill
  \begin{subfigure}[t]{.49\linewidth}
    \centering
    \includegraphics[width=\linewidth]{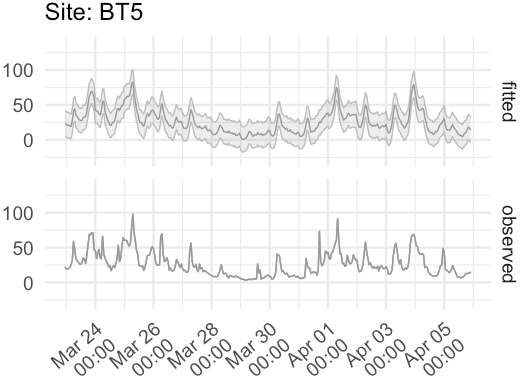}
  \end{subfigure}
\caption{Observed and fitted (with 95\% predictive bands) NO2 concentrations (in $\mu g/m^3$) at 4 different sites}
  \label{fig: fitted NO2 week12}
\end{figure*}
This section provides additional posterior predictive plots for selected monitoring sites, corresponding to Model~4 (the saturated model). These plots illustrate that the posterior predictive mean captures the overall temporal trends while remaining smoother than the observed hourly $\text{NO}_2$ concentrations.

\subsubsection{Hyper-parameter estimation by Naive Bayes}\label{apx: marg lik oprimisation}
\begin{table*}[t]
\centering
\caption{The parameter estimates obtained by maximising the log-marginal likelihood using L-BFGS algorithm provided in Stan. The time taken (in seconds) until convergence is also provided.}
\label{tab: naive bayes}
\begin{tabular}{lcccccr}
\hline
Model & $\alpha_0$ & $\alpha_1$ & $\alpha_2$ & $\alpha_3$ & $\sigma$ & Time(s)  \\ \hline
1: main & 6.86 &  5.02 & 2.20  & 0.32 & 14.9 & 3.41  \\
2: spatio-temporal interaction& 14.4 & 1.28 &  0.67 &  0.24 & 12.5 & 4.07 \\
3: all two-way interaction & 10.6 & 1.87& 1.32 &  0.31 &  8.37& 4.23 \\
4: saturated & 52.0 &  0.48 & 0.94 &  0.051 &  6.51 & 3.41 \\
5: three-way interaction only & - & 0.00079 & - & - & 36.43 & 1.36 \\
\hline
\end{tabular}
\end{table*}
In addition to obtaining samples from the posterior of the hyper-parameters, we also estimated the hyper-parameters by finding the maximiser of the log marginal likelihood \eqref{eq: log  marginal likelihood}. We used optimisation algorithm provided by Stan which can be run with the same code used for MCMC sampling. The convergence was achieved within a few seconds for all models. The values obtained (Table \ref{tab: naive bayes}) are close to those of the MCMC sample mean. 
\end{appendices}
\clearpage 

\bibliography{ref}

\begin{thebibliography}{42}
\providecommand{\natexlab}[1]{#1}
\providecommand{\url}[1]{{#1}}
\providecommand{\urlprefix}{URL }
\providecommand{\doi}[1]{\url{https://doi.org/#1}}
\providecommand{\eprint}[2][]{\url{#2}}
 \bibcommenthead

\bibitem[{Bergsma(2020)}]{bergsma2020regression}
Bergsma W (2020) Regression with {I}-priors. Econometrics and Statistics 14:89--111

\bibitem[{Bergsma and Jamil(2023)}]{bergsma2023additive}
Bergsma W, Jamil H (2023) Additive interaction modelling using {I}-priors. arXiv preprint arXiv:200715766

\bibitem[{Cheng et~al.(2019)Cheng, Ramchandran, Vatanen, Lietz{\'e}n, Lahesmaa, Vehtari, and L{\"a}hdesm{\"a}ki}]{cheng2019additive}
Cheng L, Ramchandran S, Vatanen T, et~al (2019) An additive {Gaussian} process regression model for interpretable non-parametric analysis of longitudinal data. Nature communications 10(1):1798

\bibitem[{Cooper et~al.(2022)Cooper, Martin, Hammer, Levelt, Veefkind, Lamsal, Krotkov, Brook, and McLinden}]{cooper2022global}
Cooper MJ, Martin RV, Hammer MS, et~al (2022) Global fine-scale changes in ambient {NO2} during {COVID}-19 lockdowns. Nature 601(7893):380--387

\bibitem[{Datta et~al.(2016)Datta, Banerjee, Finley, Hamm, and Schaap}]{datta2016nonseparable}
Datta A, Banerjee S, Finley AO, et~al (2016) Nonseparable dynamic nearest neighbor {Gaussian} process models for large spatio-temporal data with an application to particulate matter analysis. The annals of applied statistics 10(3):1286

\bibitem[{Durrande et~al.(2013)Durrande, Ginsbourger, Roustant, and Carraro}]{durrande2013anova}
Durrande N, Ginsbourger D, Roustant O, et~al (2013) {ANOVA} kernels and {RKHS} of zero mean functions for model-based sensitivity analysis. Journal of Multivariate Analysis 115:57--67

\bibitem[{Dutta et~al.(2021)Dutta, Kumar, and Dubey}]{dutta2021recent}
Dutta V, Kumar S, Dubey D (2021) Recent advances in satellite mapping of global air quality: evidences during {COVID}-19 pandemic. Environmental Sustainability pp 1--19

\bibitem[{Duvenaud et~al.(2011)Duvenaud, Nickisch, and Rasmussen}]{Duvenaud_2011_NIPS_additiveGP}
Duvenaud D, Nickisch H, Rasmussen C (2011) Additive {Gaussian} processes. In: Advances in Neural Information Processing Systems

\bibitem[{Duvenaud et~al.(2013)Duvenaud, Lloyd, Grosse, Tenenbaum, and Zoubin}]{duvenaud2013structure}
Duvenaud D, Lloyd J, Grosse R, et~al (2013) Structure discovery in nonparametric regression through compositional kernel search. In: International Conference on Machine Learning, pp 1166--1174

\bibitem[{Flaxman et~al.(2015)Flaxman, Wilson, Neill, Nickisch, and Smola}]{flaxman2015fast}
Flaxman S, Wilson A, Neill D, et~al (2015) Fast kronecker inference in gaussian processes with non-gaussian likelihoods. In: International conference on machine learning, PMLR, pp 607--616

\bibitem[{Gilboa et~al.(2013)Gilboa, Saat{\c{c}}i, and Cunningham}]{gilboa2013scaling}
Gilboa E, Saat{\c{c}}i Y, Cunningham JP (2013) Scaling multidimensional inference for structured gaussian processes. IEEE transactions on pattern analysis and machine intelligence 37(2):424--436

\bibitem[{Ginsbourger et~al.(2016)Ginsbourger, Roustant, Schuhmacher, Durrande, and Lenz}]{ginsbourger2016anova}
Ginsbourger D, Roustant O, Schuhmacher D, et~al (2016) On anova decompositions of kernels and gaussian random field paths. In: Monte Carlo and Quasi-Monte Carlo Methods. Springer International Publishing, Cham, pp 315--330

\bibitem[{Gronau et~al.(2020)Gronau, Singmann, and Wagenmakers}]{gronau2020bridge}
Gronau QF, Singmann H, Wagenmakers EJ (2020) {bridgesampling}: An {R} package for estimating normalizing constants. Journal of Statistical Software 92(10):1--29

\bibitem[{Groot et~al.(2014)Groot, Peters, Heskes, and Ketter}]{groot2014fast}
Groot P, Peters M, Heskes T, et~al (2014) Fast laplace approximation for {Gaussian} processes with a tensor product kernel. Proceedings of the 26th Benelux Conference on Artificial Intelligence

\bibitem[{Gu(2002)}]{gu2002smoothing}
Gu C (2002) Smoothing spline ANOVA models, vol 297. Springer

\bibitem[{Hensman et~al.(2013)Hensman, Fusi, and Lawrence}]{hensman2013gaussian}
Hensman J, Fusi N, Lawrence ND (2013) Gaussian processes for big data. In: Uncertainty in Artificial Intelligence, Citeseer, p 282

\bibitem[{Huang(1998)}]{huang1998projection}
Huang JZ (1998) Projection estimation in multiple regression with application to functional anova models. The annals of statistics 26(1):242--272

\bibitem[{Jephcote et~al.(2021)Jephcote, Hansell, Adams, and Gulliver}]{jephcote2021changes}
Jephcote C, Hansell AL, Adams K, et~al (2021) {Changes in air quality during COVID-19 ‘lockdown’in the United Kingdom}. Environmental Pollution 272:116011

\bibitem[{Kaufman and Sain(2010)}]{kaufman2010BayesianfANOVA}
Kaufman CG, Sain SR (2010) {Bayesian functional {ANOVA} modeling using Gaussian process prior distributions}. Bayesian Analysis 5(1):123 -- 149

\bibitem[{Lee et~al.(2020)Lee, Drysdale, Finch, Wilde, and Palmer}]{lee2020uk}
Lee JD, Drysdale WS, Finch DP, et~al (2020) {UK} surface $\text{NO}_2$ levels dropped by 42\% during the covid-19 lockdown: impact on surface $\text{O}_3$. Atmospheric Chemistry and Physics 20(24):15743--15759

\bibitem[{Lengerich et~al.(2020)Lengerich, Tan, Chang, Hooker, and Caruana}]{lengerich2020purifying}
Lengerich B, Tan S, Chang CH, et~al (2020) Purifying interaction effects with the functional anova: An efficient algorithm for recovering identifiable additive models. In: International Conference on Artificial Intelligence and Statistics, PMLR, pp 2402--2412

\bibitem[{Liu et~al.(2020)Liu, Ong, Shen, and Cai}]{liu2020gaussian}
Liu H, Ong YS, Shen X, et~al (2020) {When Gaussian process meets big data: A review of scalable GPs}. IEEE transactions on neural networks and learning systems 31(11):4405--4423

\bibitem[{Lu et~al.(2022)Lu, Boukouvalas, and Hensman}]{lu2022additive}
Lu X, Boukouvalas A, Hensman J (2022) Additive gaussian processes revisited. In: International Conference on Machine Learning, PMLR, pp 14358--14383

\bibitem[{M\"artens and Yau(2020)}]{pmlr-v108-martens20a}
M\"artens K, Yau C (2020) Neural decomposition: Functional anova with variational autoencoders. In: Chiappa S, Calandra R (eds) Proceedings of the Twenty Third International Conference on Artificial Intelligence and Statistics, Proceedings of Machine Learning Research, vol 108. PMLR, pp 2917--2927

\bibitem[{M{\"a}rtens et~al.(2019)M{\"a}rtens, Campbell, and Yau}]{martens2019decomposing}
M{\"a}rtens K, Campbell K, Yau C (2019) Decomposing feature-level variation with covariate gaussian process latent variable models. In: International Conference on Machine Learning, PMLR, pp 4372--4381

\bibitem[{Melkumyan and Ramos(2009)}]{melkumyan2009sparse}
Melkumyan A, Ramos FT (2009) A sparse covariance function for exact gaussian process inference in large datasets. In: Twenty-first international joint conference on artificial intelligence

\bibitem[{Nelder(1977)}]{nelder1977reformulation}
Nelder JA (1977) A reformulation of linear models. Journal of the Royal Statistical Society Series A: Statistics in Society 140(1):48--63

\bibitem[{Plate(1999)}]{plate1999accuracy}
Plate TA (1999) Accuracy versus interpretability in flexible modeling: Implementing a tradeoff using gaussian process models. Behaviormetrika 26(1):29--50

\bibitem[{Saat{\c{c}}i(2012)}]{saatcci2012scalable}
Saat{\c{c}}i Y (2012) Scalable inference for structured {Gaussian} process models. PhD thesis, University of Cambridge

\bibitem[{{Stan Development Team}(2024)}]{stan}
{Stan Development Team} (2024) Stan modeling language users guide and reference manual, version 2.35. \urlprefix\url{https://mc-stan.org}

\bibitem[{Stitson et~al.(1999)Stitson, Gammerman, Vapnik, Vovk, Watkins, and Weston}]{stitson1999support}
Stitson M, Gammerman A, Vapnik V, et~al (1999) Support vector regression with {ANOVA} decomposition kernels. {A}dvanced in kernel methods: Support vector learning

\bibitem[{Stone(1994)}]{stone1994use}
Stone CJ (1994) The use of polynomial splines and their tensor products in multivariate function estimation. The annals of statistics pp 118--171

\bibitem[{Timonen et~al.(2021)Timonen, Mannerstr{\"o}m, Vehtari, and L{\"a}hdesm{\"a}ki}]{timonen2021lgpr}
Timonen J, Mannerstr{\"o}m H, Vehtari A, et~al (2021) lgpr: an interpretable non-parametric method for inferring covariate effects from longitudinal data. Bioinformatics 37(13):1860--1867

\bibitem[{Titsias(2009)}]{titsias2009variational}
Titsias M (2009) Variational learning of inducing variables in sparse gaussian processes. In: Artificial intelligence and statistics, PMLR, pp 567--574

\bibitem[{Vehtari et~al.(2016)Vehtari, Mononen, Tolvanen, Sivula, and Winther}]{vehtari2016bayesianloo}
Vehtari A, Mononen T, Tolvanen V, et~al (2016) Bayesian leave-one-out cross-validation approximations for gaussian latent variable models. Journal of Machine Learning Research 17(103):1--38

\bibitem[{Vehtari et~al.(2017)Vehtari, Gelman, and Gabry}]{vehtari2016practical}
Vehtari A, Gelman A, Gabry J (2017) Practical bayesian model evaluation using leave-one-out cross-validation and waic. Statistics and Computing 27:1413--1432

\bibitem[{Vehtari et~al.(2024)Vehtari, Gabry, Magnusson, Yao, Bürkner, Paananen, and Gelman}]{vehtari2024loopackage}
Vehtari A, Gabry J, Magnusson M, et~al (2024) loo: Efficient leave-one-out cross-validation and waic for bayesian models. \urlprefix\url{https://mc-stan.org/loo/}, r package version 2.8.0

\bibitem[{Wahba(1990)}]{wahba1990spline}
Wahba G (1990) Spline models for observational data. Society for Industrial and Applied Mathematics, Philadelphia

\bibitem[{Williams and Seeger(2001)}]{williams2001using}
Williams C, Seeger M (2001) Using the {N}ystr{\"o}m method to speed up kernel machines. Advances in neural information processing systems 13

\bibitem[{Wilson and Nickisch(2015)}]{wilson2015kernel}
Wilson A, Nickisch H (2015) {Kernel interpolation for scalable structured Gaussian processes (KISS-GP}). In: International conference on machine learning, PMLR, pp 1775--1784

\bibitem[{Wilson et~al.(2014)Wilson, Gilboa, Nehorai, and Cunningham}]{wilson2014fast}
Wilson AG, Gilboa E, Nehorai A, et~al (2014) Fast kernel learning for multidimensional pattern extrapolation. Advances in neural information processing systems 27

\bibitem[{Yao et~al.(2018)Yao, Vehtari, Simpson, and Gelman}]{yao2018stacking}
Yao Y, Vehtari A, Simpson D, et~al (2018) {Using Stacking to Average Bayesian Predictive Distributions (with Discussion)}. Bayesian Analysis 13(3):917 -- 1007

\end{thebibliography}

\end{document}